\magnification=\magstep1
\input epsf
\voffset=0pt
\vsize=19.8 cm     
\hsize=13.5 cm
\hfuzz=2pt
\tolerance=500
\abovedisplayskip=3 mm plus6pt minus 4pt
\belowdisplayskip=3 mm plus6pt minus 4pt
\abovedisplayshortskip=0mm plus6pt
\belowdisplayshortskip=2 mm plus4pt minus 4pt
\predisplaypenalty=0
\footline={\tenrm\ifodd\pageno\hfil\folio\else\folio\hfil\fi}

\def\la{\mathrel{\hbox{\rlap{\hbox{\lower4pt\hbox{$\sim$}}}\hbox{$<$}}}}
\def\ga{\mathrel{\hbox{\rlap{\hbox{\lower4pt\hbox{$\sim$}}}\hbox{$>$}}}}

\def\arcmin{\hbox{$^\prime$}}

\def\utw{\smash{\rlap{\lower5pt\hbox{$\sim$}}}}
\def\udtw{\smash{\rlap{\lower6pt\hbox{$\approx$}}}}

\def\getsto{\mathrel{\hbox{\rlap{$\gets$}\hbox{\raise2pt\hbox{$\to$}}}}}
\def\lid{\mathrel{\hbox{\rlap{\hbox{\lower4pt\hbox{$=$}}}\hbox{$<$}}}}
\def\gid{\mathrel{\hbox{\rlap{\hbox{\lower4pt\hbox{$=$}}}\hbox{$>$}}}}
\def\sol{\mathrel{\hbox{\rlap{\hbox{\raise4pt\hbox{$\sim$}}}\hbox{$<$}}}
}
\def\sog{\mathrel{\hbox{\rlap{\hbox{\raise4pt\hbox{$\sim$}}}\hbox{$>$}}}
}
\def\lse{\mathrel{\hbox{\rlap{\hbox{\raise4pt\hbox{$<$}}}\hbox{$\simeq$}
}}}
\def\gse{\mathrel{\hbox{\rlap{\hbox{\raise4pt\hbox{$>$}}}\hbox{$\simeq$}
}}}
\def\grole{\mathrel{\hbox{\lower2pt\hbox{$<$}}\kern-8pt
\hbox{\raise2pt\hbox{$>$}}}}
\def\leogr{\mathrel{\hbox{\lower2pt\hbox{$>$}}\kern-8pt
\hbox{\raise2pt\hbox{$<$}}}}
\def\loa{\mathrel{\hbox{\rlap{\hbox{\lower4pt\hbox{$\approx$}}}\hbox{$<$
}}}}
\def\goa{\mathrel{\hbox{\rlap{\hbox{\lower4pt\hbox{$\approx$}}}\hbox{$>$
}}}}

%
%

\font\kleinhalbcurs=cmmib10 scaled 833
\font\eightrm=cmr8
\font\sixrm=cmr6
\font\eighti=cmmi8
\font\sixi=cmmi6
\skewchar\eighti='177 \skewchar\sixi='177
\font\eightsy=cmsy8
\font\sixsy=cmsy6
\skewchar\eightsy='60 \skewchar\sixsy='60
\font\eightbf=cmbx8
\font\sixbf=cmbx6
\font\eighttt=cmtt8
\hyphenchar\eighttt=-1
\font\eightsl=cmsl8
\font\eightit=cmti8

\font\bxf=cmbx10
  \mathchardef\Gamma="0100
  \mathchardef\Delta="0101
  \mathchardef\Theta="0102
  \mathchardef\Lambda="0103
  \mathchardef\Xi="0104
  \mathchardef\Pi="0105
  \mathchardef\Sigma="0106
  \mathchardef\Upsilon="0107
  \mathchardef\Phi="0108
  \mathchardef\Psi="0109
  \mathchardef\Omega="010A
\def\rahmen#1{\vskip#1truecm}
\def\begfig#1cm#2\endfig{\par
\setbox1=\vbox{\rahmen{#1}#2}%
\dimen0=\ht1\advance\dimen0by\dp1\advance\dimen0by5\baselineskip
\advance\dimen0by0.4true cm
\ifdim\dimen0>\vsize\pageinsert\box1\vfill\endinsert
\else
\dimen0=\pagetotal\ifdim\dimen0<\pagegoal
\advance\dimen0by\ht1\advance\dimen0by\dp1\advance\dimen0by1.4true cm
\ifdim\dimen0>\vsize
\topinsert\box1\endinsert
\else\vskip1true cm\box1\vskip4true mm\fi
\else\vskip1true cm\box1\vskip4true mm\fi\fi}
\def\figure#1#2{\smallskip\setbox0=\vbox{\noindent\petit{\bf Fig.\ts#1.\
}\ignorespaces #2\smallskip
\count255=0\global\advance\count255by\prevgraf}%
\ifnum\count255>1\box0\else
\centerline{\petit{\bf Fig.\ts#1.\ }\ignorespaces#2}\smallskip\fi}

\def\xfigure#1#2#3#4{\midinsert\noindent
    $$\epsfxsize=#4truecm\epsffile{#3}$$
    \figure{#1}{#2}\endinsert}


\def\begtab#1cm#2\endtab{\par
\ifvoid\topins\midinsert\vbox{#2\rahmen{#1}}\endinsert
\else\topinsert\vbox{#2\kern#1true cm}\endinsert\fi}
\def\rahmen#1{\vskip#1truecm}
\def\begpet{\vskip6pt\bgroup\petit}
\def\endpet{\vskip6pt\egroup}
\def\begref{\par\bgroup\petit
\let\it=\rm\let\bf=\rm\let\sl=\rm\let\INS=N}
\def\petit{\def\rm{\fam0\eightrm}%
\textfont0=\eightrm \scriptfont0=\sixrm \scriptscriptfont0=\fiverm
 \textfont1=\eighti \scriptfont1=\sixi \scriptscriptfont1=\fivei
 \textfont2=\eightsy \scriptfont2=\sixsy \scriptscriptfont2=\fivesy
 \def\it{\fam\itfam\eightit}%
 \textfont\itfam=\eightit
 \def\sl{\fam\slfam\eightsl}%
 \textfont\slfam=\eightsl
 \def\bf{\fam\bffam\eightbf}%
 \textfont\bffam=\eightbf \scriptfont\bffam=\sixbf
 \scriptscriptfont\bffam=\fivebf
 \def\tt{\fam\ttfam\eighttt}%
 \textfont\ttfam=\eighttt
 \normalbaselineskip=9pt
 \setbox\strutbox=\hbox{\vrule height7pt depth2pt width0pt}%
 \normalbaselines\rm
\def\vec##1{\setbox0=\hbox{$##1$}\hbox{\hbox
to0pt{\copy0\hss}\kern0.45pt\box0}}}%
\let\ts=\thinspace
%
\font \tafontt=     cmbx10 scaled\magstep2
\font \tafonts=     cmbx7  scaled\magstep2
\font \tafontss=     cmbx5  scaled\magstep2
\font \tamt= cmmib10 scaled\magstep2
\font \tams= cmmib10 scaled\magstep1
\font \tamss= cmmib10
\font \tast= cmsy10 scaled\magstep2
\font \tass= cmsy7  scaled\magstep2
\font \tasss= cmsy5  scaled\magstep2
\font \tasyt= cmex10 scaled\magstep2
\font \tasys= cmex10 scaled\magstep1
\font \tbfontt=     cmbx10 scaled\magstep1
\font \tbfonts=     cmbx7  scaled\magstep1
\font \tbfontss=     cmbx5  scaled\magstep1
\font \tbst= cmsy10 scaled\magstep1
\font \tbss= cmsy7  scaled\magstep1
\font \tbsss= cmsy5  scaled\magstep1

\newbox\chsta\newbox\chstb\newbox\chstc
\def\centerpar#1{{\advance\hsize by-2\parindent
\rightskip=0pt plus 4em
\leftskip=0pt plus 4em
\parindent=0pt\setbox\chsta=\vbox{#1}%
\global\setbox\chstb=\vbox{\unvbox\chsta
\setbox\chstc=\lastbox
\line{\hfill\unhbox\chstc\unskip\unskip\unpenalty\hfill}}}%
\leftline{\kern\parindent\box\chstb}}
 \def \chap#1{
    \vskip24pt plus 6pt minus 4pt
    \bgroup
 \textfont0=\tafontt \scriptfont0=\tafonts \scriptscriptfont0=\tafontss
 \textfont1=\tamt \scriptfont1=\tams \scriptscriptfont1=\tamss
 \textfont2=\tast \scriptfont2=\tass \scriptscriptfont2=\tasss
 \textfont3=\tasyt \scriptfont3=\tasys \scriptscriptfont3=\tenex
     \baselineskip=18pt
     \lineskip=18pt
     \raggedright
     \pretolerance=10000
     \noindent
     \tafontt
     \ignorespaces#1\vskip7true mm plus6pt minus 4pt
     \egroup\noindent\ignorespaces}%
 \def \sec#1{
     \vskip25true pt plus4pt minus4pt
     \bgroup
 \textfont0=\tbfontt \scriptfont0=\tbfonts \scriptscriptfont0=\tbfontss
 \textfont1=\tams \scriptfont1=\tamss \scriptscriptfont1=\kleinhalbcurs
 \textfont2=\tbst \scriptfont2=\tbss \scriptscriptfont2=\tbsss
 \textfont3=\tasys \scriptfont3=\tenex \scriptscriptfont3=\tenex
     \baselineskip=16pt
     \lineskip=16pt
     \raggedright
     \pretolerance=10000
     \noindent
     \tbfontt
     \ignorespaces #1
     \vskip12true pt plus4pt minus4pt\egroup\noindent\ignorespaces}%
 \def \subs#1{
     \vskip15true pt plus 4pt minus4pt
     \bgroup
     \bxf
     \noindent
     \raggedright
     \pretolerance=10000
     \ignorespaces #1
     \vskip6true pt plus4pt minus4pt\egroup
     \noindent\ignorespaces}%
 \def \subsubs#1{
     \vskip15true pt plus 4pt minus 4pt
     \bgroup
     \bf
     \noindent
     \ignorespaces #1\unskip.\ \egroup
     \ignorespaces}
 \def \subsub#1{
     \vskip15true pt plus 4pt minus 4pt
     \bgroup
     \bf
     \noindent
     \ignorespaces #1\unskip\ \egroup
     \ignorespaces}
\def\footnoterule{\kern-3pt\hrule width 2true cm\kern2.6pt}
\newcount\footcount \footcount=0
\def\advftncnt{\advance\footcount by1\global\footcount=\footcount}
\def\fonote#1{\advftncnt$^{\the\footcount}$\begingroup\petit
       \def\textindent##1{\hang\noindent\hbox
       to\parindent{##1\hss}\ignorespaces}%
\vfootnote{$^{\the\footcount}$}{#1}\endgroup}

\newcount\sterne
\outer\def\byebye{\bigskip\typeset
\sterne=1\ifx\speciali\undefined\else
\bigskip Special caracters created by the author
\loop\smallskip\noindent special character No\number\sterne:
\csname special\romannumeral\sterne\endcsname
\advance\sterne by 1\global\sterne=\sterne
\ifnum\sterne<11\repeat\fi
\vfill\supereject\end}
\def\typeset{\centerline{\petit This article was processed by the author
using the \TeX\ Macropackage from Springer-Verlag.}}

\def\eck#1{\left\lbrack #1 \right\rbrack}
\def\eckk#1{\bigl[ #1 \bigr]}
\def\rund#1{\left( #1 \right)}
\def\abs#1{\left\vert #1 \right\vert}

\def\ave#1{\left\langle #1 \right\rangle}

\def\part#1#2{{\partial #1\over\partial #2}}

{\catcode`\@=11
\gdef\SchlangeUnter#1#2{\lower2pt\vbox{\baselineskip 0pt \lineskip0pt
  \ialign{$\m@th#1\hfil##\hfil$\crcr#2\crcr\sim\crcr}}}
}
\def\gtrsim{\mathrel{\mathpalette\SchlangeUnter>}}
\def\lesssim{\mathrel{\mathpalette\SchlangeUnter<}}
\def\Re{{\cal R}\hbox{e}}

\def\D{{\cal D}}
\def\I{{\cal I}}

\def\i{{\rm i}}
\def\d{{\rm d}}

\def\Real{{\rm I\mathchoice{\kern-0.70mm}{\kern-0.70mm}{\kern-0.65mm}%
  {\kern-0.50mm}R}}
\def\C{\rm C\kern-.42em\vrule width.03em height.58em depth-.02em
       \kern.4em}
\font \bolditalics = cmmib10
\def \vc #1{{\textfont1=\bolditalics \hbox{$\bf#1$}}}
{\catcode`\@=11
\def\malen#1#2{\global \advance\fnum by 1\midinsert\vskip0.5truecm\noindent%
$${\epsfxsize=0.5\hsize\epsffile{#1}}$$
     $$\vbox{\hsize=158truemm{
        {\eightpoint\par\noindent
    {\bf Figure \the\fnum:~}#2}}}$$\medskip\noindent\endinsert}
\def\ma#1#2{\global \advance\fnum by 1\vskip0.5truecm\noindent%
$${\epsfxsize=0.6\hsize\epsffile{#1}}$$
     $$\vbox{\hsize=158truemm{
        {\eightpoint\par\noindent
    {\bf Figure \the\fnum:~}#2}}}$$\medskip\noindent}

\def\eps{{\epsilon}}

\def\U{{\cal U}}
\chap{Steps towards nonlinear cluster inversion through gravitational 
distortions: \hfill \break
III. Including a redshift distribution of the sources}
{\bf Carolin Seitz \& Peter Schneider}
\bigskip\noindent
Max-Planck-Institut f\"ur Astrophysik, Postfach 1523, D-85740
Garching, Germany.

\sec{Abstract}
In a series of previous papers we have considered the reconstruction
of the surface mass density of a cluster of galaxies from images of
lensed faint background galaxies. We
showed that the reconstructed surface mass density is not uniquely
determined, but that there exists a global invariance transformation
that leaves the shape of the images of the lensed galaxies unchanged. Because of
this, only lower limits on the total mass of a cluster can be derived
if no further informations besides image ellipticities are used.
Throughout these papers we used the
simplifying assumption that all sources are at the same redshift.

In this paper we account for a redshift distribution of the faint
galaxies, and in particular, some of these galaxies can lie in front of
the cluster or can be cluster members. We show how the mass
distribution of a cluster of galaxies can be obtained from images of
these faint galaxies, if the redshift
distribution of these galaxies is known. We demonstrate that for the
reconstruction of non-critical clusters we need less information on the
redshift distribution of the galaxies, i.e., we only need to know two
or three moments of the distribution.

We show that the mean mass density across the data field is still a
free variable, i.e., there remains a global invariance transformation
of the resulting mass density field. For non-critical clusters we can
derive the transformation explicitly and it is similar to that derived
previously for the case of a single redshift of the sources. We discuss
several theoretical ideas to break the mass degeneracy; of those
considered, we find that
only the magnification effect on the number density of galaxy
images can be used successfully in practice.

\vfill \eject

\sec{1 Introduction}
This is the third and final paper in a series in which we have
considered the reconstruction of the surface mass density of galaxy
clusters from the (weak) image distortion it imposes onto faint
background galaxies. Inspired by the pioneering work of Kaiser \&
Squires (1993, hereafter KS), who derived a parameter-free inversion
equation for the surface mass density of the cluster in terms of the
tidal deflection field [which has been discovered earlier by Tyson,Valdes
\& Wenk (1990); earlier work on cluster mass determination by weak
lensing effects include Kochanek (1990) and Miralda-Escud\'e (1991)]
we have started to generalize the KS inversion method to include also
strong clusters, i.e., cluster which are -- at least nearly -- capable
to produce giant arcs. In Paper\ts I (Schneider \& Seitz 1995) we have
analyzed the basic observable from image distortions and pointed out a
global invariance transformation of the surface mass density which
leaves the observable distortion invariant. In Paper\ts II (Seitz \&
Schneider 1995) we have then developed an iterative procedure to
reconstruct the density field, which was then applied to synthetic
data and shown to work well.

In the present paper, we want to generalize the treatment of Paper\ts
II in two different ways. First, the inversion procedure constructed
in Paper\ts II is not unbiased and contains boundary artefacts in the
same way as the original KS method, due to the fact that the
observations are always limited to a finite data field.  There are now
several different inversion methods which are unbiased (Schneider
1995; Kaiser et al.\ 1995; Bartelmann 1995, Seitz \& Schneider 1996) 
-- all of them are based on a relation found by Kaiser
(1995). As was demonstrated in Seitz \& Schneider (1996), 
these finite-field methods work
well and can be applied efficiently. Second, in the earlier papers
cited above it was assumed that all sources have the same effective
redshift; by that we mean that all sources have about the same ratio
$D(z_{\rm d},z)/ D(z)$ of the angular diameter distance as measured
from the lens and from the observer. This assumption is fairly well
justified if the cluster redshift is relatively small, $z_{\rm
d}\lesssim 0.2$, say.  For weak cluster lenses, this assumption can be
easily dropped since then the (linear) reconstruction proceeds
equivalently to the case that all sources are at about the `mean
redshift' of the population (this will be made more precise in
Sect.\ts 4.3 below). However, if the cluster is not assumed to be
weak, the consideration of the source redshift distribution becomes
more difficult. We shall discuss this problem in some detail below.

The rest of the paper is organized as
follows: in Sect.\ts 2 we present basic equations and our notation.
In Sect.\ts3 we derive the dependence of the local observables 
on the local values of the surface mass density and the
shear. In Sect.\ts 4, we describe the nonlinear reconstruction
method; after briefly reviewing the unbiased finite-field inversion
method developed by Seitz \& Schneider (1996) for the case of a 
single source redshift, we
generalize this technique to the case of a redshift distribution. As
in the case of a single source redshift, we also have a global invariance
transformation, which, however, cannot be written in closed form in
general. However, restricting the consideration to moderately strong
clusters, we construct the invariance transformation explicitly. 
As we show in Sect.\ts 5 and in Appendix\ts3, the presence of a
redshift distribution of the sources
 provides several methods to break the invariance transformation. 

However, from simulations we found that among the methods considered,
only that is successful in practice which makes use of the
magnification effect on the number density of galaxy images (see also
Broadhurst, Taylor \& Peacock 1995).  In contrast to Broadhurst,
Taylor \& Peacock (1995) and Broadhurst (1995) we do not use the local
magnification to derive the local mass density, but we use the
magnification averaged over the field to constrain the mean mass
density in the field (in order to combine information from the
shear with local magnification information, a maximum likelihood
approach seems to be the best strategy -- see Bartelmann et al.\ts
1996). Essential for the success of this method is
that the source counts deviate sufficiently strongly from the
$n(S)\propto S^{-2}$ behaviour, which seems to be the case for
galaxies with red colour.  Furthermore, the method depends strongly on the
assumption that the faint background galaxies are distributed rather
smoothly, i.e., that no strong correlations in their angular position
is present, as seems to be justified by recent investigations (e.g., Infante
\& Pritchet 1995).

We apply our
methods to synthetic data in Sect.\ts 6 to demonstrate its
feasibility. The application to an HST exposure of the cluster
Cl0939+4713 will be published elsewhere (Seitz et al. 1995c). 
We discuss our results in Sect.\ts 7. 

\sec{2 Basic equations}
\subs{2.1 Redshift dependence of the lens equation}
The distortion of images of background galaxies depends on the
dimensionless surface mass density of the lens, which is the physical
surface mass density $\Sigma(\vc\theta)$ divided by the critical
surface mass density $\Sigma_{\rm crit}$. If we now consider the
background sources to be distributed in redshift, then the critical
surface mass density depends on the redshift $z$ of the source:
$$
\Sigma_{\rm crit}(z)=
\cases{\displaystyle \qquad \;
\infty  \qquad \qquad  \qquad \hbox{for } \quad z \le z_{\rm d} ,
 \cr
\displaystyle {c^2 D(z)\over 4\pi G D(z_{\rm d}) D(z,z_{\rm d})}\
  \qquad \hbox{for } \quad z > z_{\rm d} . \cr}
\eqno(2.1)
$$
Here $D(z_{\rm d})$ and $D(z)$ are the angular
diameter-distances from the observer to the lens at redshift $z_{\rm
d}$ and to the source at
redshift $z$,
and $D(z_{\rm d},z)$ is the angular diameter-distance from the lens to the
source. 
Defining 
$$
w(z;z_{\rm d})=
{\lim_{z\to \infty} \Sigma_{\rm crit}(z) \over  \Sigma_{\rm
crit}(z)}={\Sigma_{\rm crit_\infty}\over \Sigma_{\rm crit}(z)}
\eqno(2.2)
$$
we obtain for the dimensionless surface mass density
$\kappa(\vc\theta,z)$ at angular position $\vc\theta$ for a source at
redshift $z$ 
$$
\kappa(\vc \theta,z)={\Sigma( \vc \theta)
\over \Sigma_{\rm crit}(z)}={\Sigma
(\vc \theta)
\over \Sigma_{\rm crit_\infty}}
{\Sigma_{\rm crit_\infty}
\over \Sigma_{\rm crit}(z)}\equiv \kappa(\vc \theta) \; w(z;z_{\rm d})
\ . 
\eqno(2.3)
$$
The function $w(z;z_{\rm d})$ relates the `lensing strength' for a
source with redshift $z$ to that of a hypothetical source at `infinite
redshift', 
and its form depends on the geometry of the
universe. For an Einstein--de Sitter universe we have
$$
w(z;z_{\rm d})=\cases
{\displaystyle {\qquad 0 \quad \qquad \qquad \qquad{\rm for}\quad
{z\le z_{\rm d}}} \cr
\displaystyle \cr
\displaystyle{
{\sqrt{1+z}-\sqrt{1+z_{\rm d}} \over \sqrt{1+z}-1}\qquad {\rm for}
\quad {z > z_{\rm d}}\; .}\cr}
 \eqno(2.4)
$$
In particular, for sources with redshift smaller than that of the
lens, the `lensing strength' vanishes. For the rest of this paper, we
consider a single cluster lens at redshift $z_{\rm d}$, and drop the
second argument of $w$, i.e., $w(z;z_{\rm d})\equiv w(z)$.

Since the shear $\gamma$ is related linearly to the surface mass
density, its dependence on source redshift is the same as for $\kappa$,
$$
\gamma(\vc\theta,z)=\gamma_1(\vc \theta,z)+{\rm i} \gamma_2(\vc
\theta,z) = {1\over \pi}\int_{\Real^2}\d^2\theta'\; 
\D(\vc\theta-\vc\theta')\,\kappa(\vc\theta',z)\equiv w(z)\;
\gamma(\vc \theta) \quad,
\eqno (2.5)
$$
where the complex kernel is $\D(\vc x)=(x_1^2-x_2^2+2{\rm i}x_1x_2)/\abs{x}^4$.
Because of the relations (2.3) \& (2.5) it is sufficient to consider
the dimensionless surface mass density and shear with respect to one
particular source redshift, which is at infinity for our choice. 

The lens equation for a source with redshift $z$ is
$$\vc \beta=\vc \theta-\nabla \psi(\vc \theta,z)\ , 
\eqno(2.6)
$$
where $\psi(\vc\theta,z)=w(z)\,\psi(\vc\theta)$ is the deflection
potential at $\vc\theta$  
for a source at redshift $z$. Hence, the linearized lens equation at
$\vc\theta$ which describes the image distortion of small sources is
given by
$$ 
A(\vc \theta,z)\equiv{\partial \vc\beta\over\partial\vc\theta}=
\I -w(z)
\left ( \matrix{ \psi_{11} & \psi_{12}\cr
\psi_{21} & \psi_{22} \cr } \right )
\eqno (2.7)
$$
where $\I$ is the two-dimensional unit matrix and $\psi_{ij}={\partial
^2 \psi\over \partial \theta_i\partial \theta_j }$. The magnification
$\mu(\vc \theta)$
of an image with position $\vc \theta$ and a source redshift $z$ then
becomes
$$
\mu(\vc\theta,z):={1\over \abs{\det
A(\vc\theta,z)}}={1\over \abs{[1-w(z)\kappa(\vc\theta)]^{2}
-w^2(z)\abs{\gamma(\vc\theta)}^2}}\quad .
\eqno (2.8)
$$
The cluster is non-critical for sources at
redshift $z$ if $\det A(\vc\theta,z)>0$ everywhere; it is
non-critical for all source redshifts if
$[1-\kappa(\vc \theta)]^2-\abs{\gamma(\vc \theta)}^2>0$ for all $\vc\theta$.

%
\sec{3 Local observables and their dependence on local lens
parameters $\kappa$ and $\gamma$}
Throughout this section we assume that these lensing
parameters can be considered as constant over a small solid angle
around $\vc\theta$, and we will therefore suppress the argument
$\vc\theta$ in all equations.

In Papers\ts I\ts\&\ts II we described the shape of an image by the complex
number
$$
\chi={Q_{11}-Q_{22}+2\i Q_{12}\over Q_{11}+Q_{22}}\ ,
\eqno(3.1)
$$
where $Q_{ij}$ are the components of the tensor of second moments of
the surface brightness of the image. The same relation was used to
define the source ellipticity $\chi_{\rm s}$ in terms of the second
brightness moments $Q_{\rm s}$ of the source.
An image with elliptical isophotes
has $\abs \chi=(1-r^2)/(1+r^2)$, if $r\in[0,1]$ is the axis
ratio of the ellipse. Here, we define 
$$
\eps={\chi\over 1+\sqrt{1-\abs{\chi}^2}} 
\eqno(3.2)
$$
as the ellipticity parameter;
for an image with elliptical isophotes, $\abs{\eps}=(1-r)/(1+r)$, and
the phase of $\eps$ is the same as that of $\chi$.
Defining
$$
g(z)= {\gamma(z)\over 1-\kappa(z)}={w(z)\gamma\over 1-w(z)
\kappa}\; ,
\eqno(3.3)
$$
the transformation between source
ellipticity $\chi_{\rm s}$ and image ellipticity $\chi$ for 
a source with redshift $z$ has been shown in Paper\ts I to be
$$
\chi={\chi_{\rm s}-2 g(z)+g^2(z)\chi_{\rm s}^*   \over
1+\abs{g(z)}^2-2 \Re\eckk{g(z)\chi_{\rm s}^*}}\quad ,
\eqno (3.4a)
$$
where an asterisk denotes complex conjugation. 
From this transformation we find with (3.2) the transformation between
the ellipticity parameter $\eps$ of an image and that of the
corresponding source $\eps_{\rm s}$:
$$
\eps(\eps_{\rm s},z)=\cases
{\displaystyle {\eps_{\rm s}-g(z) \over 1-g^*(z)\; \eps_{\rm s}} \quad {\rm for}
\quad {\abs{g(z)} \le 1\; ,} \cr
\displaystyle \cr
\displaystyle { 1-g(z)\; \eps_{\rm s}^* \over \eps_{\rm s}^*-g^*(z) }
\; \quad {\rm for} \quad {\abs {g(z)} >1\; .} \cr}
\eqno(3.4b)
$$
The condition $\abs{g(z)}\le 1$ ($\abs{g(z)} >1$) is equivalent to the
condition $\det A(z)\ge 0$ ($\det A(z)<0$). 

Now, let $p_{\eps_{\rm s}}(y)\, y\; \d y\, \d\varphi$ be the probability that the
source ellipticity $\eps_{\rm s}=y\; \exp(2{\rm i}\varphi)$ is within
$y\,\d y\,\d\varphi$ around $\eps_{\rm s}$. Then, 
for fixed redshift $z$, the expectation value of the $n$-th moment
$\eps^n$ is given through
$$
\ave{\eps^n}_{\eps_{\rm s}}(z)
\equiv\int_0^1 \;\d y\;y\, p_{\eps_{\rm s}}(y)\int_0^{2\pi} \; \d \varphi\;
\eps^n(\eps_{\rm s},z)=\cases
{\displaystyle \eckk{-g(z)}^n\qquad {\rm for}
\quad {\abs{g(z)} \le 1\; ,}\cr
\displaystyle \rund { - 1\over g^*(z)}^n 
\; \quad {\rm for} \quad {\abs {g(z)} >1\; .} \cr }
\eqno(3.5)
$$
The remarkable fact that the expectation values
$\ave{\eps^n}_{\eps_{\rm s}}$ do not depend on the source
ellipticity distribution, whereas $\ave{\chi}_{\chi_{\rm s}} $ does
(Papers\ts I\ts \&\ts II), is the reason for choosing $\eps$ as the
ellipticity parameter in this paper. We derive (3.5) in the Appendix\ts1.

If the galaxies are distributed in redshift
according to the probability density $p_z(z)$, the expectation
values of the moments $\eps^n$ become:
$$\eqalign{
&\ave{\eps^n}_{\eps_{\rm s},z}\equiv \int_0^\infty  \d z\; p_z(z)\;
\ave{\eps^n}_{\eps_{\rm s}}(z)\cr
&= 
\int_{\det A(z)\ge 0} \d z \; p_z(z)\eckk{-g(z)}^n 
+\int_{\det A(z)< 0} \d z \; p_z(z)\rund {-1\over g^*(z)}^n 
\cr
&= \gamma^n \int_{\det A(w)\ge 0} \d w\;  p_w(w) \rund{- w\over
1-\kappa w}^n + 
\rund{1\over \gamma^*}^n \int_{\det A(w)<0 } \d w \; p_w(w) \rund{
1-\kappa w\over  - w}^n  
\cr
&\equiv  \gamma^n\; X_n(\kappa,\gamma) +{1\over \gamma^{*n}}
Y_n(\kappa,\gamma)= 
\gamma^n\rund{X_n(\kappa,\gamma)+ {1\over
\abs{\gamma}^{2n}}Y_n(\kappa,\gamma)} . \cr } 
\eqno(3.6)
$$
In the third line we used the transformation of the source redshift
distribution into their $w$-distribution, which consists of a delta
`function' at $w=0$ for galaxies with $z\le z_{\rm d}$ and is given by
$p_w(w)\,\d w=p_z(z)\,\d z$ for galaxies with $z>z_{\rm d}$. For a
single redshift $z_0$ of the sources, the expectation values reduce to 
$\ave{\eps^n}_{\eps_{\rm s},z}=\ave{\eps^n}_{\eps_{\rm s}}(z_0)$ given
in (3.5). 

Generally, the boundaries of the integrals in (3.6), and therefore
$X_n$ and $Y_n$, depend on $\kappa$ and $\gamma$,
$$
\eqalign{
X_n(\kappa,\gamma)&=\rund{\int_0^{\min(1,1/[\kappa+\abs{\gamma}])}
+\int_{1\over \max(1,\kappa-\abs{\gamma})}^1} \d w\;  p_w(w) \rund{- w\over
1-\kappa w}^n \cr
Y_n(\kappa,\gamma)&=\int_{\min(1,1/[\kappa+\abs{\gamma}])}
^{1\over \max(1,\kappa-\abs{\gamma})} \d w \; p_w(w) \rund{
1-\kappa w\over  - w}^n  \quad .\cr }
\eqno (3.7)
$$
One sees that 
for some values of the parameters $\kappa $ and
$\gamma$ there exist two intervals of $w$ for which 
$\det A(w)\ge 0$, 
seperated by that interval for which $\det A(w)<0$. 

In the case of weak lensing we obtain from (3.6) that $\ave{\eps^n}_{\eps_{\rm
s},z}\approx(-1)^n\ave{w^n}\gamma^n$, and the shear is -- modulo
$\ave{w^n}$ -- an observable.

An estimate of these expectation values for the moments of the image
ellipticities can be obtained by considering (locally) an ensemble of
sources with ellipticities $\eps_k$ and defining the means
$$
\bar{\eps^n}\equiv{\sum_{k=1}^{N_{\rm gal}}\eps^n_k \; u_k\over 
\sum_{k=1}^{N_{\rm gal}} u_k} \ ,
\eqno(3.8)
$$
where $u_k$ is an appropriately chosen weight factor (see Paper\ts II
or Seitz \& Schneider 1996).
These mean values $\bar{\eps ^n}$ are statistically
distributed around the expectation values $\ave{\eps^n}_{\eps_{\rm
s},z}$, and we use them as an estimate 
for
$\ave{\eps^n}_{\eps_{\rm s},z}$.

To summarize, the mean image ellipticity and their higher moments
$\ave{\eps^n}_{\eps_{\rm s},z}$ do no longer directly provide us with
a local estimate for $\gamma/(1-\kappa)$ or its inverse as in the case
where all sources are at the same redshift, which was assumed in
previous papers.  In fact, the dependence of $\ave{\eps}_{\eps_{\rm
s},z}$ on the local lens parameters $\kappa$ and $\gamma$ can be quite
complicated, depending on the redshift distribution of the sources and
on the local lens parameters itself.

To obtain information on the local lens parameters $\kappa$, $\gamma$
from $\bar \eps\approx \ave{\eps}_{\eps_{\rm s},z}$ we have to know
the redshift distribution $p_z(z)$ of the galaxy population which
enters Eq.\ts(3.6).  We point out that the average in (3.6) extends
over all galaxies, i.e., also over those situated in front of the
lens, which, however, do not provide any information on the local lens
parameters, since $\ave{\eps^n}_{\eps_{\rm s}}(z)=0$ for $z\le z_{\rm
d}$, but contribute to the noise which is inherent in the estimate of
the local lens parameters derived from $\bar \eps$.  In the rest of
this paper we assume that this redshift distribution of galaxies is
known or at least can be estimated, e.g., by the lensing effect itself
(see Bartelmann \& Narayan 1995, Kneib et al. 1995).

For illustration we assume that the redshift distribution is given
by a function of the form
$$
p_z(z)={\beta z^2\over \Gamma\rund{3/\beta}z_0^3}
\; \exp\rund{-\rund{z/z_0}^\beta}
\ , 
\eqno(3.9)
$$
taken from Brainerd et al. (1995),
with moments
$\ave z=3 \ts z_0$ for $\beta=1$, $\ave z=1.5\ts z_0$ for $\beta=1.5$
and $\ave z=\Gamma(4/3)\ts z_0$ for $\beta=3$.

In Fig.\ts1 we show this distribution $p(z)$ for $\beta=1=\ave z$ and the
corresponding distribution $p_w(w)$ for a lens with
redshift $z_{\rm d}=0.4$.
\xfigure{1}{The lower panel (solid line) shows the redshift
distribution given in Eq.\ts(3.9) for $\ave z=3 z_0=1$ and $\beta=1$.
The dashed line shows the function $w(z)$ defined in Eq.\ts(2.4) for a
cluster redshift of $z_{\rm d}=0.4$. The upper panel shows the
corresponding distribution $p_w(w)$. Note that $p_w(w)$ has a
delta-`function' peak at $w=0$, with an amplitude given by the
probability that a source has a redshift smaller than that of
the lens}
{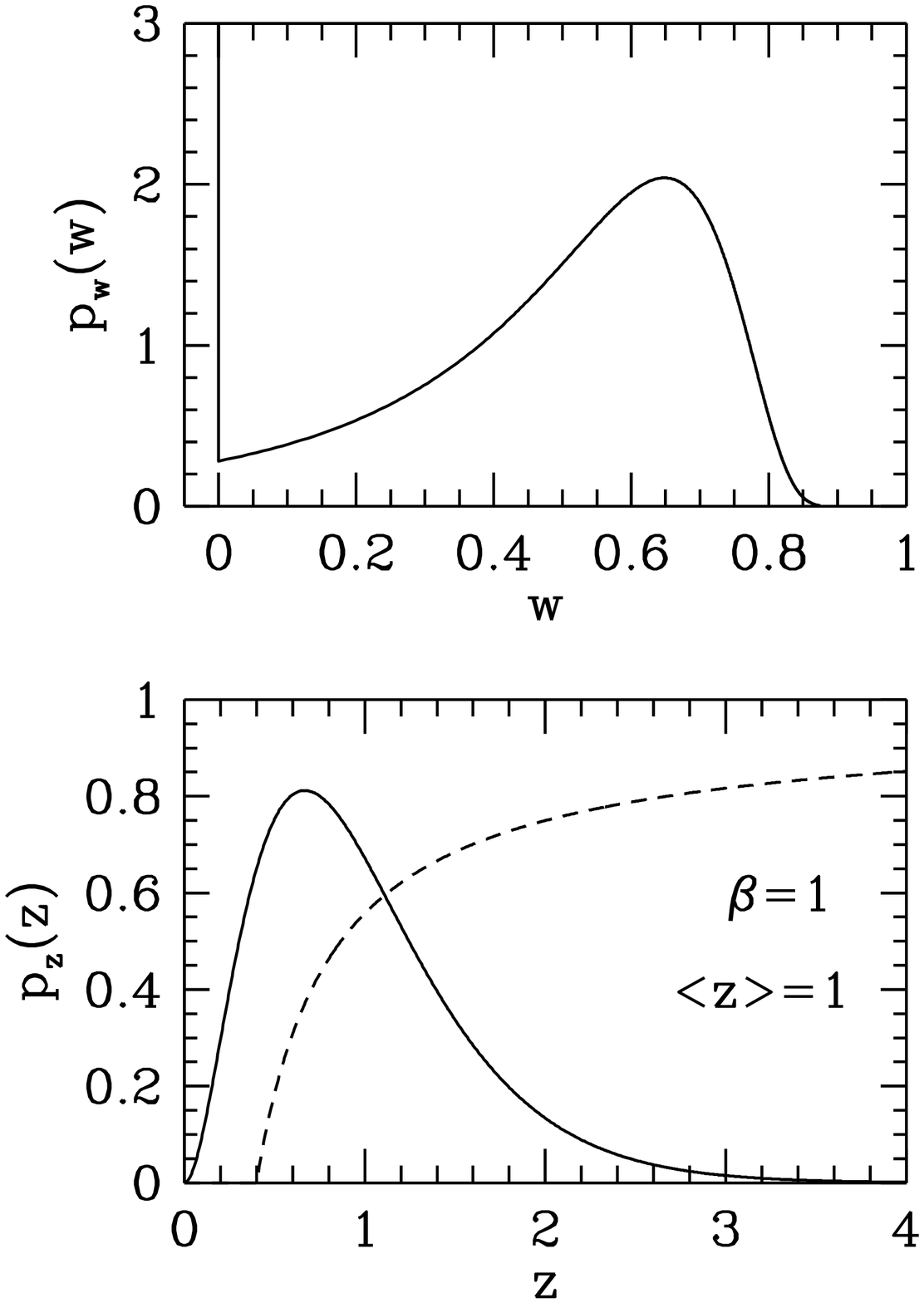}{8}
\sec{4  The reconstruction of the cluster mass distribution}
%

\subs{4.1 Inversion relations for a single source redshift}
The reconstruction of the surface mass density of a cluster from
lensed images of background sources has been described in several
papers (see introduction) for sources at the same redshift $z$. Let
$\kappa(\vc\theta)$ be the dimensionless surface mass density of the
cluster, scaled by the appropriate critical surface mass density, and
let $\gamma(\vc\theta)$ be the corresponding complex shear. Both
quantities are given as second partial derivatives of the deflection
potential, and it was shown by Kaiser (1995) that the following
relation between third partial derivatives of the deflection
potential, or first partial derivatives of $\kappa$ and $\gamma$, is
valid: 
$$
\vc \nabla \kappa(\vc\theta)=-\left ( \matrix{ \gamma_{1,1}+\gamma_{2,2} \cr
\gamma_{2,1}-\gamma_{1,2}  \cr } \right )\equiv \vc U(\vc\theta)
\eqno(4.1)
$$
Hence, the surface mass density can be obtained in terms of the shear
by integrating this first-order differential equation. 
However, the shear is not an observable in general (Kochanek 1990; Paper\ts
I), but the quantity
$g(\vc\theta)=\gamma(\vc\theta)/\eckk{1-\kappa(\vc\theta)}$, if we
confine our considerations to non-critical clusters. 
(For critical clusters only $2g/(1+\abs
g^2)$ is observable.) Inserting
$\gamma=g\rund{1-\kappa}$ into (4.1) yields (Kaiser 1995)
$$
\nabla K(\vc\theta)={1\over 1-g_1^2-g_2^2}
\pmatrix{1+g_1 & g_2\cr g_2 & 1-g_1\cr}\pmatrix{g_{1,1}+g_{2,2}\cr
g_{2,1}-g_{1,2}\cr }\equiv \vc u(\vc\theta)\quad,
\eqno (4.2)
$$
where
$$
K(\vc\theta):=\ln \eck{1-\kappa(\vc\theta)}\quad .
\eqno (4.3)
$$
Hence, it is possible to derive the gradient of the quantity $K$ in
terms of the observable quantity $g$. Obviously, the integration of
(4.2) allows an arbitrary integration constant, i.e., $K(\vc\theta)$
can only be determined up to an additive constant. Several methods
exist to perform an integration of (4.2); see Schneider (1995), Kaiser et
al. (1995), Bartelmann (1995), and Seitz \& Schneider (1996). All of them would be equivalent if
$\vc u(\vc\theta)$ were a gradient vector field. However, since $\vc
u$ is determined observationally, it is not exact and thus in general
not rotation-free. The integration derived in Seitz \& Schneider (1996),
$$
K(\vc\theta)=\int_\U \d^2\theta'\;\vc H(\vc \theta';\vc \theta)\cdot
\vc u(\vc \theta') + \bar K \quad ,
\eqno (4.4)
$$
was explicitly constructed to account for this `rotational noise
component' and has been demonstrated in Seitz \& Schneider (1996) to work better than the
other proposed methods. Here, $\vc H(\vc \theta';\vc \theta)$ is a
vector field explicitly constructed in Seitz \& Schneider (1996), and $\bar K$ is the average
of $K(\vc\theta)$ over the data field $\U$, i.e., the region where
image ellipticities have been measured. Of course, $\bar K$ is an
undetermined constant, so that the surface mass density
$\kappa(\vc\theta)$ is determined up to the transformation
$$
\kappa(\vc\theta) \to (1-\lambda) + \lambda \kappa(\vc\theta)\quad .
\eqno (4.5)
$$

\subs{4.2 General inversion method}
In the general case of a redshift distribution of the sources, we
again make use of (4.1). The formal integration of (4.1) proceeds in 
the same way as that of (4.2), i.e., 
$$
\kappa(\vc\theta)=\int_\U \d^2\theta'\;\vc H(\vc\theta';\vc \theta)
\cdot \vc U(\vc\theta') + \bar\kappa\quad ,
\eqno (4.6)
$$
where $\bar\kappa$ is the average of $\kappa$ over the data field
$\U$. The vector field $\vc U(\vc\theta)$ is defined in (4.1) and
given by first partial derivatives of the shear. The shear in turn is
related to the mean image ellipticity via (3.6),
$$
\ave{\eps}_{\eps_{\rm s},z}=\gamma\rund{X_1(\kappa,\gamma)+{1\over \abs{\gamma}^2}
Y_1(\kappa,\gamma)}\approx {\bar \eps} \quad .
\eqno (4.7)
$$
Note that Eq.\ts (4.7) is a local
relation, valid at every point $\vc\theta$. The complicated dependence
of $\ave{\eps}_{\eps_{\rm s},z}$ on $\kappa$ and $\gamma$ suggests an iterative
approach for the solution of the inversion problem: let $\bar \eps$ be
an `measured' estimate for $\ave \eps _{\eps_{\rm s},z}$ and let
$\gamma^{(n)}(\vc\theta)$ and $\kappa^{(n)}(\vc\theta)$ be an estimate
for the shear field and the surface mass density. From that, an
updated estimate for the shear field can be obtained, using (4.7):
$$
\gamma^{(n+1)}(\vc\theta)={\bar \eps}
\eck{ X_1\rund{\kappa^{(n)}(\vc\theta),\gamma^{(n)}(\vc\theta)}
+{Y_1\rund{\kappa^{(n)}(\vc\theta),\gamma^{(n)}(\vc\theta)} \over
\abs{\gamma^{(n)}(\vc\theta)}^2}}^{-1}\quad .
\eqno (4.8)
$$
Then, by
differentiation, the vector field $\vc U^{(n+1)}(\vc\theta)$ can be
calculated from (4.1), by using the shear field
$\gamma^{(n+1)}(\vc\theta)$. And finally, an updated estimate for the
surface mass density field is obtained from (4.6),
$$
\kappa^{(n+1)}(\vc\theta)=\int_\U \d^2\theta'\;\vc H(\vc\theta';\vc \theta)
\cdot \vc U^{(n+1)}(\vc\theta') + \bar\kappa\quad .
\eqno (4.9)
$$
This iteration process is started by chosing
$\kappa^{(0)}(\vc\theta)=0$, $\gamma^{(0)}(\vc\theta)=0$. It is clear
that the integration constant $\bar \kappa$ is still a free variable,
i.e., with the method described here there remains a global invariance
transformation of the resulting surface mass density field; in
contrast to the case considered in Sect.\ts 4.1, this transformation
cannot be explicitly determined, due to the highly nonlinear relations
occurring here. 
For critical clusters we need at most 10 steps to achieve a
convergence of the iteration algorithm. For less massive clusters
about $5$ iteration steps are sufficient. We find that the iteration
algorithm is more stable and converges faster than in the case of a
single source redshift (see Paper\ts II), mainly because there are no
well-defined critical curves as function of $\vc\theta$, since their
location depends on the source redshift.

\subs{4.3 The case for non-critical clusters}
If the cluster is non-critical for all redshifts of the sources,
the inversion problem can be simplified because
 $Y_n(\kappa,\gamma)=0$ for all $n$ -- see (3.7)
-- and the $X_n$ depend only on $\kappa$, 
$$
X_n(\kappa)=\int_0^1\d w\;p_w(w)\,\rund{-w\over 1-\kappa w}^n\quad .
\eqno (4.10)
$$
Then, (3.6) simplifies to
$$
\ave{\eps^n}_{\eps_{\rm s},z}=\gamma^n\,X_n(\kappa)\ .
\eqno (4.11)
$$
As a result, the iteration
procedure described in the last subsection can be applied in a
somewhat simpler way, by using (4.11) instead of (4.7); in
addition, one can use the approximation (A2.4) derived in the
Appendix\ts 2 for $X_1(\kappa)$ which yields
$$
X_1(\kappa) \approx {-\ave w\over 1-\kappa {\ave{w^2} \over
\ave w}} \ .
\eqno(4.12)
$$
This approximation is found to be
sufficiently accurate to describe $X_1(\kappa)$ for (generic) non-critical
clusters (see
Fig.\ts8 in Appendix\ts2).
 
Combining (4.11) for $n=1$, (4.12), and replacing 
the expectation value $\ave{\eps}_{\eps_{\rm s},z}$ with the
observed local average $\bar\eps$, we obtain
$$
\gamma=-{\bar\eps\over \ave w}\ts 
\rund{1- f \ave w \kappa}\quad ,
\eqno (4.13)
$$
with the definition 
$$
f:={\ave{w^2}\over \ave{w}^2}\quad .
\eqno (4.14)
$$
Inserting this expression for
$\gamma$ into (4.1), one obtains after some manipulations
$$
V\,\nabla\kappa={1\over \ave{w}}\rund{1-f\ave{w} \kappa}
\pmatrix{\bar\eps_{1,1}+\bar\eps_{2,2} \cr
\bar\eps_{2,1}-\bar\eps_{1,2} \cr}\quad ,
\eqno (4.15)
$$
with the matrix
$$
V=\pmatrix{1+f\bar \eps_1 & f\bar\eps_2 \cr
             f\bar\eps_2 & 1-f\bar\eps_1 \cr }\quad .
$$
Thus, 
$$
\nabla\kappa={1\over \ave w}\rund{1-f\ave w\kappa}\;V^{-1}\,
\pmatrix{\bar\eps_{1,1}+\bar\eps_{2,2} \cr
\bar\eps_{2,1}-\bar\eps_{1,2} \cr}\quad ,
\eqno (4.16)
$$
with the inverse
$$
V^{-1}={1\over 1 - f^2({\bar\eps_1}^2+{\bar\eps_2}^2)}
\pmatrix{1-f\bar \eps_1 & -f\bar\eps_2 \cr
            - f\bar\eps_2 & 1+f\bar\eps_1 \cr }\quad .
\eqno (4.17)
$$
If we now define
$$
K(\vc\theta):=\ln\rund{1-f\ave w\kappa(\vc\theta)}\quad ,
\eqno (4.18)
$$
(4.16) can be written as
$$
\nabla K(\vc\theta)= f\,V^{-1}\,\pmatrix{\bar\eps_{1,1}+\bar\eps_{2,2} \cr
\bar\eps_{2,1}-\bar\eps_{1,2} \cr}\equiv \vc u(\vc\theta)\quad .
\eqno (4.19)
$$
Thus, the vector field $\vc u(\vc\theta)$ can be constructed directly
from the observable $\bar\eps$, as in the case of all sources being at
the same redshift, and the same inversion equation (4.4) should be
used, but with the current definitions of $K$ and $\vc u$. Thus, by
using the approximation (4.12), the inversion of a non-critical
cluster is no more complicated than in the case of a single source redshift.
In particular, (4.18) immediately shows that $K$ can be
determined only up to an additive constant, which implies the
invariance transformation
$$
\kappa(\vc\theta) \to {(1-\lambda)\over f\ave{w}} +\lambda
\kappa(\vc\theta) \quad ;
\eqno (4.20)
$$
in other words, $\rund{1-\ave{w^2} \kappa(\vc\theta)/\ave{w}}$ can be
determined only up to a multiplicative constant.

\sec{5 Breaking the mass invariance}
In Sect.\ts 4, we have shown that there exists an invariance
transformation for the surface mass density map which leaves the
observed local mean image ellipticities constant (4.5, 4.9 \& 4.20). 
The occurrence of
the invariance can be traced back to the fact that the inversion
equation was derived from a first-order differential equation -- see
(4.1) -- which has a free integration constant.
We now discuss several ideas to determine this constant.

\subs{5.1 Using only shape information}
\subsubs{5.1.1 A maximum-likelihood approach}
We consider a fixed set of
`observed'  galaxies with ellipticities $\eps_i$ and positions $\vc
\theta_i$. Then we reconstruct the surface mass density of a
cluster as described in Sect.\ts 4.2 assuming a value $\bar \kappa$
for the mean mass density in the field [see Eq.\ts(4.9)]. From that we
calculate the mass density $\kappa(\vc \theta_i)$ 
and shear $\gamma(\vc \theta_i) $ 
at the positions $\vc \theta_i$ ($i=1,...,N$). Next we calculate the
probability $p(\eps_i,\kappa(\vc \theta_i),\gamma(\vc \theta_i))$ 
to observe the image ellipticity $\eps_i$ at the galaxy
position $\vc \theta_i$. Finally we maximize the likelihood ${\cal L}_1=\Pi_i
 p(\eps_i,\kappa(\vc \theta_i),\gamma(\vc \theta_i))$ using different
values for $\bar \kappa$ in the mass reconstruction.
We find that the mass degeneracy is broken, but unfortunately only
weakly, unless the
redshift distribution is broad and the surface mass density is
unrealistically high. From various simulations we conclude that the
likelihood of observing a set of galaxy ellipticities $\eps(\vc
\theta_i)$ gives no significant limits on the mean mass density in the
field. Furthermore, the likelihood depends much stronger on the
assumed redshift distribution -- which in practice is weakly
constrained -- than on the value of the mean mass density $\bar \kappa$
and is therefore a much better tool to investigate the redshift
distribution of the sources than to determine the mean surface mass
density of a particular cluster.

\subsubs{5.1.2 Second moments of image ellipticities}
For non-critical clusters, 
the ratio 
$$
R={\abs{\ave{\eps^2}_{\eps_{\rm
s},z}}\over \abs{\ave{\eps}_{\eps_{\rm s},z}}^2}={X_2(\kappa)\over
X_1(\kappa)^2}\approx f \; { \rund{1-\kappa {\ave{w^2}\over \ave w}}^2
\over \rund{1-\kappa {\ave {w^3}\over \ave{w^2}} }^2}  
\eqno(5.1)
$$ 
depends only on the local surface mass density $\kappa$. We derive
(5.1) in Appendix\ts3.  Measuring $R$ could therefore provide a direct
estimate for $\kappa$. However, we can not measure
${\ave{\eps^2}_{\eps_{\rm s},z}}$ and ${\ave{\eps}_{\eps_{\rm s},z}} $
accurately enough to determine the local mass density $\kappa$ or the
average mass density $\bar
\kappa$ from a resonable number density of 
image ellipticities (see Appendix\ts3).

\subs{5.2 Magnification information through number density of images}
Let us assume that the number density $n_0(S,z)$ of the unlensed faint
galaxies with flux $S$ and redshift $z$ is given through
$$
n_0(S,z)=p_z(z)\; F(S) \ ,
\eqno(5.2)
$$
where $p_z(z)$ is the normalized redshift distribution and $F(S)$ the
distribution in flux. Note that the factorization of $n_0(S,z)$ as
given in (5.2) is a fairly special assumption for the galaxy
distribution, which will not be valid in general. However, for
simplicity we shall make use of (5.2) in this paper. This
factorization may in fact be an approximate distribution over a limited
range of flux $S$. Since the magnifications are not very large except
perhaps in the very central parts of the cluster, the `dynamic range'
over which (5.2) is applied is not large and might be a valid
description. It should be noted that this factorization has already
been used implicitly in (3.6). If (5.2) is not assumed, then the
redshift distribution of sources locally -- where the surface mass
density is $\kappa$ and the shear is $\gamma$  -- will differ from
$p_z(z)$ due to the magnification, so that the functions $X_n$ and
$Y_n$ will attain an additional dependence on $\kappa$ and
$\abs{\gamma}$. This will not cause any additional conceptual problem.

Then, consider a position $\vc \theta$ in
the cluster with surface mass density $\kappa(\vc \theta)$ and shear
$\gamma(\vc \theta)$.  The observed number density of galaxies with
redshift $z$ and flux larger than $S$ is (see Broadhurst et al. 1995)
$$
n(>S,z,\vc \theta)=p_z(z)\; {1\over \mu(\vc \theta,z) }\; F\rund{>{S\over
\mu(\vc \theta,z)}} \ ,
\eqno(5.3)
$$
with the magnification defined in (2.8)
The total number density of galaxies observed
with flux larger than $S$ is obtained through integration of (5.3) and
yields
$$
n(>S,\vc \theta)=\int_0^\infty \d z\; p_z(z)\; {1\over \mu(\vc \theta,z)
}\;F\rund{>{S\over \mu(\vc \theta,z)}} \ .
\eqno(5.4)
$$
If $F(>S)\propto S^{-\alpha}$, then we obtain from (5.4)
$$
n(>S,\vc \theta)=n_0(>S) \int_0^\infty \d z\;p_z(z)\; \mu^{\alpha-1}(\vc
\theta, z)\equiv n_0(>S)\ave{\mu^{\alpha-1}(\vc \theta,z)}_z\ .
\eqno(5.5)
$$
From Eq.\ts(5.5) we conclude that the number density is not changed
if $\alpha=1$ and it is increased (decreased) for $\alpha>1$
($\alpha<1$).  Next, averaging (5.5) over the data field $\cal U$
with area $ U$, we obtain
$$
\ave { n(>S,\vc \theta)}_{\cal U}\equiv
{1\over U} \int_{\cal U}\d^2 \theta \; n(>S,\vc \theta) =n_0(>S)
\ave{\mu^{\alpha-1} (\vc \theta,z)}_{z,{\cal U}} \ .
\eqno(5.6)
$$
Hence, the ratio of the number of observed galaxies in the data field
$\cal U$ to the number $U\; n_0(>S)$ which would be observed in the
absence of a foreground lens gives $\ave{\mu^{\alpha-1} (\vc
\theta,z)}_{z,{\cal U}} $. The number density $n_0(>S)$ of the unlensed
galaxies is regarded as an universal function and has been measured in
several colours down to very faint magnitudes (see for example
Smail et al.\ts 1995). Because of this, the local observables in the case
of cluster lensing are the moments of the image ellipticities
$\ave{\eps^n}(\vc \theta)$ and $\ave{\mu(\vc \theta,z)}_z$. We note
that the result that $\ave{\mu(\vc \theta,z)}_z$ is a local observable
is only true for the ansatz (5.2) with $F(>S)\propto S^{-\alpha}$. If
either the ansatz (5.2) does not hold, or if $F(>S)$ is not a power
law, then the observable quantity is a different one and can in
particular have a much more complicated dependence on the
magnification. However, $n(>S,\vc\theta)$ can still be compared 
-- may be not analytically -- with $n_0(>S)$ and provides information 
on the local magnification.

In the following we consider a fixed value of the flux threshold $S$
and use for notational simplicity
$n(\vc \theta)$ and $n_0$ instead of  $n(>S,\vc \theta)$ and
$n_0(>S)$.

\subsub{5.2.1 Do galaxy positions break the mass
invariance?}
The probability to observe a galaxy within the area $\d^2\theta$
around $\vc \theta$ is 
$$
p(\vc \theta)\,\d^2\theta={n(\vc \theta)\over \int_{\cal U}\d^2\theta
\;n(\vc\theta)} \d^2\theta\ .
\eqno(5.7)
$$
Using Eqs.\ts(5.5) and (5.6) we obtain this probability in terms of the local
(redshift-averaged) magnification 
$$
p(\vc \theta)\,\d^2\theta={{1\over U} \ave{\mu^{\alpha-1}(\vc \theta,z)}_z
\over \ave{\mu^{\alpha-1}(\vc \theta,z)}_{z,{\cal U}}} \;
\d^2\theta \ .
\eqno(5.8)
$$
The likelihood ${\cal L}$ for observing $N $ galaxies with
positions $\vc \theta_i$ ($i=1,\dots,N$) is
$$
{\cal L}=\Pi _{i}\;  p(\vc \theta_i)\ .
\eqno(5.9)
$$
On the other hand, we can perform the mass reconstruction according to
Eq.\ts(4.9) and obtain $\ave{\mu^{\alpha-1}(\vc \theta,z,\bar
\kappa)}_z $ as a function of the assumed mean mass density $\bar \kappa$ in the
field. From that we can calculate with Eq.\ts (5.8) the probability
density $p(\vc\theta_i)$ and finally according to (5.9) the
likelihood ${\cal L} (\bar \kappa)$ for the assumed mean mass
density.  However this does not work. To understand this, assume for a
moment that all galaxies are at the same redshift; for simplicity we
take this redshift to be `infinity' here. Then the reconstructed mass
density $\kappa$ is related to the true one $\kappa_{\rm true}$
through the invariance transformation (4.5) which gives
$(1-\kappa_{\rm true})\lambda=1-\kappa$ and $\gamma_{\rm
true}\lambda=\gamma$. For the magnification we obtain
$$
\mu^{\alpha-1}_{\rm true}(\vc \theta)=
\mu^{\alpha-1}(\vc \theta)\lambda^{2-2\alpha} \ ,
\eqno(5.10)
$$
and from that we conclude that $p(\vc \theta)$ is independent of
$\lambda$, or, equivalently, independent of $\bar \kappa$.  This means
that for all sources at the same redshift, the {\it positions} of the observed
galaxies do not give any information on the {\it mean mass density} in
the field. 

If the sources are distributed in redshift then the mean
mass $\bar \kappa$ assumed for the reconstruction changes the ratio 
$$
p(\vc \theta,\bar \kappa)={
{1\over U} \ave{\mu^{\alpha-1}(\vc \theta,z,\bar \kappa)}_z
\over \ave{\mu^{\alpha-1}(\vc \theta,z,\bar \kappa)}_{z,{\cal U}}}\ .
\eqno(5.11)
$$
However, the dependence on $\bar \kappa$ is very weak as can be
verified for non-critical clusters through calculating, for fixed
$\kappa(\vc \theta)$ and $\gamma(\vc
\theta)$, the function 
$\ave{\mu^{\alpha-1}(\vc \theta,z)}_z(\lambda)$ using the invariance
transformation (4.20).
As a result, the {\it positions} of the observed galaxies provide no
or not enough information to break the mass degeneracy. 

Up to here all our attempts to break the mass degeneracy in practice
have failed. 
This is because for all cases considered, the
mass degeneracy is -- in theory -- broken because of the presence of a
redshift distribution of the sources and is no longer broken if all
sources are at the same redshift.  We conclude that in order to
derive $\bar \kappa$ one has to use a method which breaks the mass
degeneracy not because of the presence of a redshift distribution of
the sources, but independent of that.

\subsubs{5.2.2 Breaking the mass degeneracy through the total number of
observed galaxies}
We shall now consider the magnification effect on the distribution of
galaxy images. We do not use the magnification effect locally as
proposed by Broadhurst et al. (1995) and Broadhurst (1995) in order
to derive the local mass 
density from it, but we use the `global' magnification effect, i.e.,
we compare the total number of observed galaxies with the expected
(unlensed one), and relate that to the mean mass density inside the
observed field.

In the above subsection we have not used the number density $n_0$ of
galaxies observable in the absence of lensing. 
If we use this, then we can calculate from Eq.(5.6) 
the expected number $\ave
N(\bar \kappa)$ of galaxies as a function of the mean mass density
$\bar \kappa$ assumed in the mass reconstruction,
$$
\ave N(\bar \kappa)=U \; n_0 \ave{\mu^{\alpha-1}(\vc
\theta,z,\bar \kappa)}_{z,{\cal U}} \ 
\eqno(5.12) 
$$ 
and compare that with the number $N$ of galaxies actually observed.
We compare the deviation in terms of the standard deviation
$\sigma=\sqrt{\ave N}$ and obtain
$$
\chi^2(\bar \kappa)={\eck{N-\ave N (\bar \kappa)}^2\over \ave N (\bar \kappa)}
\ .
\eqno(5.13)
$$
The value $\bar \kappa$ for which $\chi^2(\bar \kappa)
$ shows a minimum (in fact, at which $\chi^2=0$) gives the most 
probable value for the mean surface
mass density of the cluster. 

We note that in the case of a single redshift of the sources (see
Sect.\ts 4.1) the determination of the mean mass density is even
simpler: the mass reconstruction is done according to (4.4) with an
assumed value for $\bar K$, say $\bar K=0$. From the resulting mass
and shear maps $\kappa(\vc \theta)$ and $\gamma(\vc \theta)$, one can
calculate $\ave{\mu^{\alpha-1}(\vc \theta)}_{{\cal U}}$. Because
of Eq.\ts(5.6) we can relate that to the true magnification and from
Eq.\ts(5.10) we conclude that
$$
{N\over U n_0}
={\ave{n(\vc \theta)}_{\cal U}\over n_0}
=\ave{\mu_{\rm true}^{\alpha-1}(\vc \theta)}_{{\cal U}}
=\lambda^{2-2\alpha}\ave{\mu^{\alpha-1}(\vc \theta)}_{{\cal U}} \ .
\eqno(5.14)
$$
Thus, we can immediately calculate $\lambda$ from (5.14), and 
with the invariance transformation (4.5) the true mass density
$\eck{1-\kappa_{\rm true}(\vc\theta)}\lambda=1-\kappa(\vc\theta)$. 

In Sect.\ts 6.2, we apply this method to break the mass invariance to
a numerically simulated cluster ($z_{\rm d}=0.4$), for the case that
the sources are distributed in redshift.  We demonstrate that the
method yields good estimates for the mean surface mass density, i.e.,
it yields good estimates on the total mass inside the observed
field. However, we point out that the method is not applicable in the
case that $\alpha\approx 1$, because then no magnification bias or
antibias occurs, and the observed number density of galaxy images
provides no information on the mean mass density $\bar \kappa$ in the
field.

We note that the method suggested by 
Bartelmann \& Narayan (1995) still works in this case:
they made use of the magnification
of individual galaxy images, at fixed surface brightness which is
known to be unchanged by gravitational light deflection. Their method
will become extremely useful once sufficient HST data become available
to obtain the mean size-surface brightness relation for these faint
galaxies.

\sec{6 Application to simulated data}
For testing the reconstruction technique we use the same numerically
modelled cluster as in Paper II. Using its surface mass density shown
in Fig.\ts 2,  we generate a distribution of
synthetic galaxy images using the source
ellipticity distribution 
$$
p_{\rm s}\rund{\abs{\eps_{\rm s}}}={1\over \pi R^2\eck{1-\exp\rund{-1/R^2}}}
\; \exp\rund{-\abs{\eps_{\rm s}}^2/R^2} \quad {\rm with} \quad R=0.15\ ,
\eqno(6.1)
$$
and the redshift-distribution (3.9) with $\beta_{\rm true}=1$ and
$\ave z_{\rm true}=1$. 
The galaxies are randomly distributed in the source plane, and the
number density of galaxy images is $n_{\rm
gal}=60 /{\rm arcmin}^2$. The cluster is placed at a redshift
$z_\d=0.4$ so that $\ave w=0.44$,
$\ave{w^2}=0.25 $ and $f=\ave{w^2}/\ave w^2=1.29$.
\subs{6.1 The derived mass distribution for different assumed redshift
distributions of the sources}
We show in Fig.\ts 3 the resulting surface mass density map for $\bar
\kappa=\bar\kappa_{\rm true}=0.12$, $\beta=\beta_{\rm true}=1$ and $\ave
z=\ave z _{\rm true}=1$, i.e., the assumed redshift distribution is
the true one.  We find a good agreement with the original mass density
displayed in Fig.\ts2. Compared to the mass reconstruction shown Paper
II -- where all sources were at the same redshift -- the noise is
increased due to the dependence of the lensing strength of the cluster
on the redshift of the sources.
 
However, in practice we do not know $\bar \kappa_{\rm true}$ and we
therefore choose $\bar \kappa$ such that the minimum of the -- locally
smoothed -- reconstructed surface mass density is zero. The reason for
not requiring the minimum of the reconstructed mass map to be zero but
that of the locally smoothed mass density (gaussian smoothing function with
smoothing length of $1\arcmin$) is that the former quantity is much more
affected by noise. Choosing $\bar \kappa$ in that way leads to a good
reconstruction only if the minimum of the cluster mass distribution is
indeed close to zero, what is valid in practice if the observed field
is several Mpc wide.

With that procedure we obtain almost the same mass density as that
shown in Fig.\ts 3 and the mean mass density detected inside the field is
$\bar \kappa=0.133 $. 

\xfigure{2}{The surface mass density distribution of the numerically
modelled cluster. The sidelength is about $10'$ corresponding to
3.8\ts Mpc for $H_0=50$\ts km/s/Mpc and an EdS universe. 
The levels of the contourlines are
$0.1,0.2,\dots,1.3$.}
{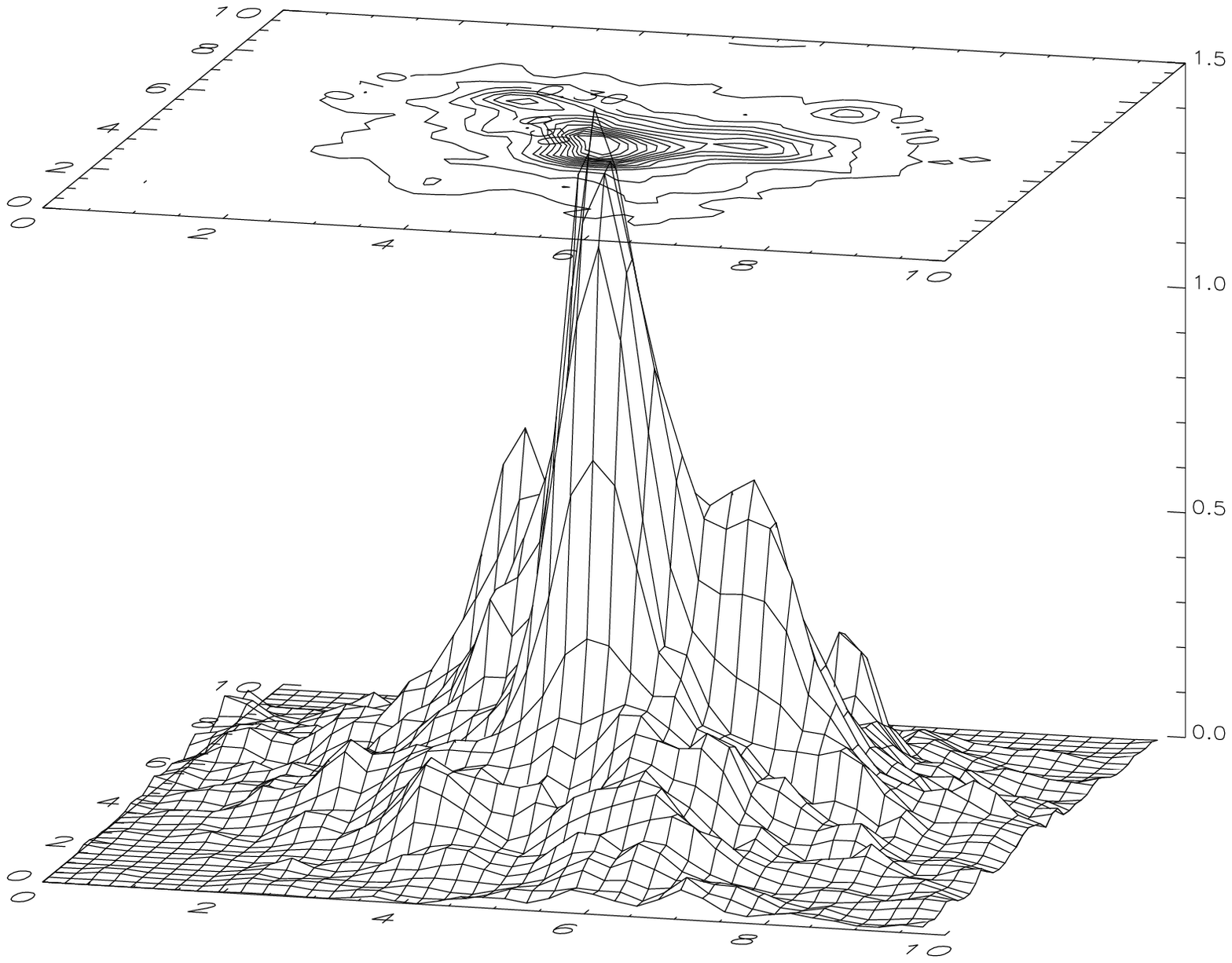}{10}

\xfigure{3}{The reconstructed mass density obtained solving
Eq.\ts(4.9) iteratively, assuming that the redshift distribution is
known and assuming $\bar\kappa=\bar \kappa_{\rm true}=0.12$. We obtain
almost the same result if we choose $\bar \kappa$ such that the
minimum of the locally-smoothed mass density is zero. This
assumption leads to $\bar \kappa=0.133$. The levels of the
contourlines are the same as in Fig.\ts 2}{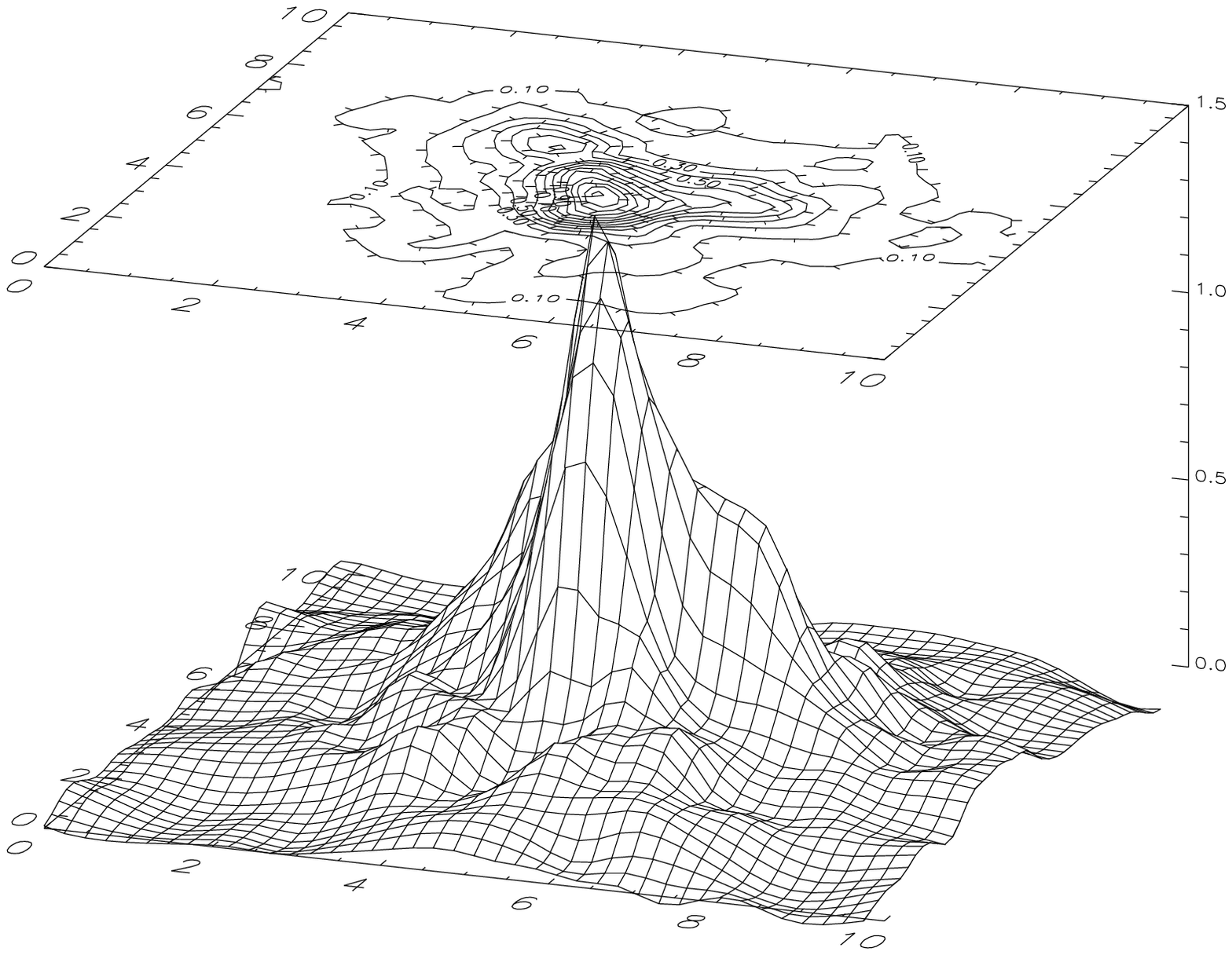}{10}

In Fig.\ts 4 we show the mass distribution obtained using $\ave z=1.5$
and $\beta=2$. The mean mass density $\bar\kappa =0.098$ is chosen
such that the minimum of the locally-smoothed mass density is
zero. Overestimating the mean redshift of the galaxies leads to an
underestimation of the overall mass distribution and vice versa. This
is shown in Fig.\ts 5, where we use $\ave z=0.6$ and $\beta=3$ for the
reconstruction and obtain a mean mass density of $\bar \kappa=0.228$.
Common to the Figs.\ts 3 to 5 is that the substructure is nicely
recovered, regardless of the assumed redshift distribution.

In Fig.\ts 6 we show the mean mass density obtained for different
assumed redshift distributions. We find that this is approximately
proportional to $1/\ave w$, i.e., it is inversely proportional to
the assumed mean effective lensing strength. 

\xfigure{4}{The reconstructed mass density obtained solving
Eq.\ts(4.9) iteratively assuming that the redshift distribution given
through (3.9) with $\ave z=1.5 $ and $\beta=2$. The true redshift
distribution is described by $\ave z_{\rm true}=1$ and $\beta_{\rm true}=1$.
For the mean mass density we obtain $\bar \kappa=0.098$. The levels of the
contourlines are the same as in Fig.\ts 2}{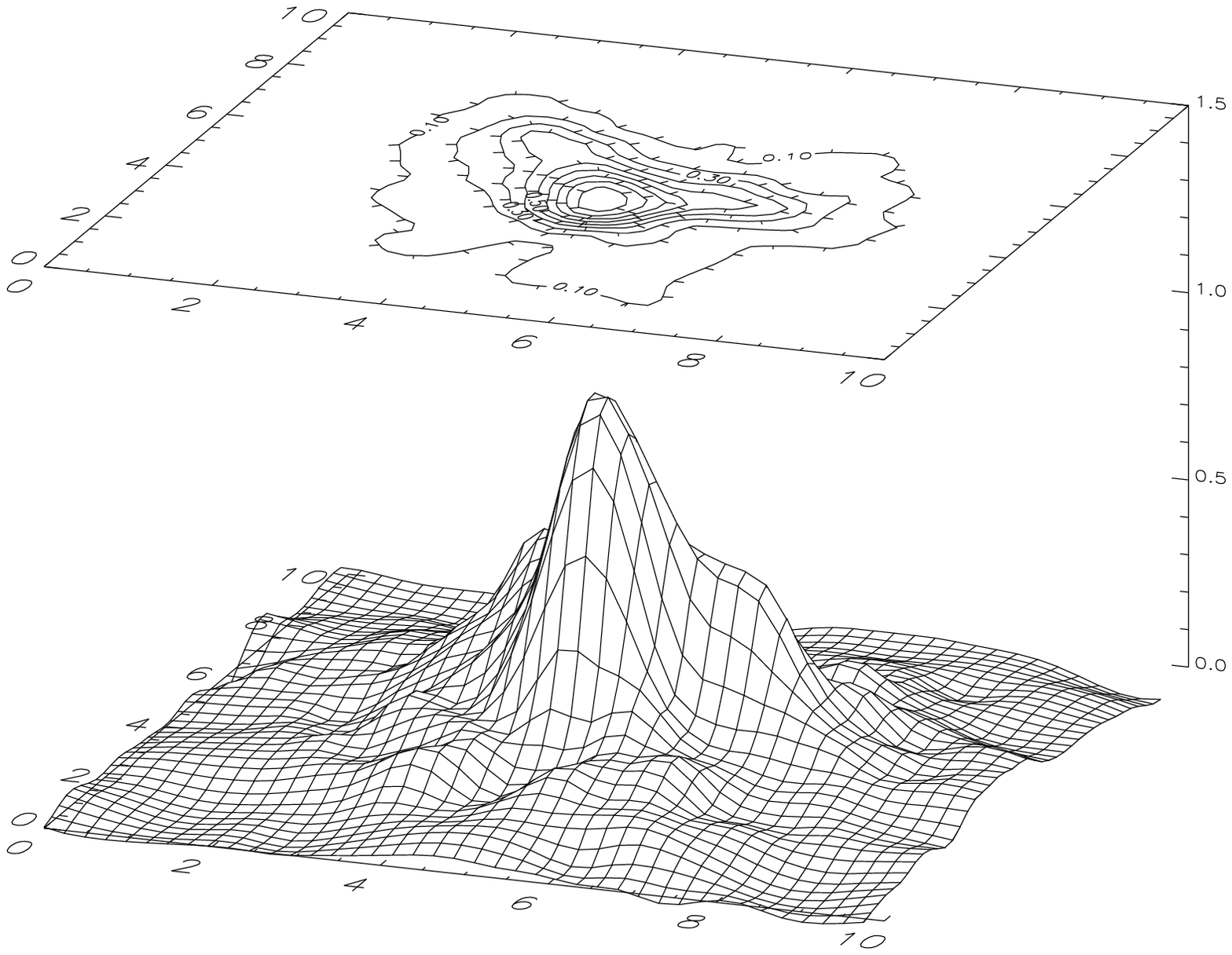}{10}

\xfigure{5}{Same as Fig.\ts 4 but assuming that the 
redshift distribution is given
through (3.9) with $\ave z=0.6 $ and $\beta=3$.  For the mean mass
density we obtain $\bar \kappa=0.228$. The levels of the contourlines
are the same as in Fig.\ts 2. Note that the scale of the $z$-axis is
increased in this figure}{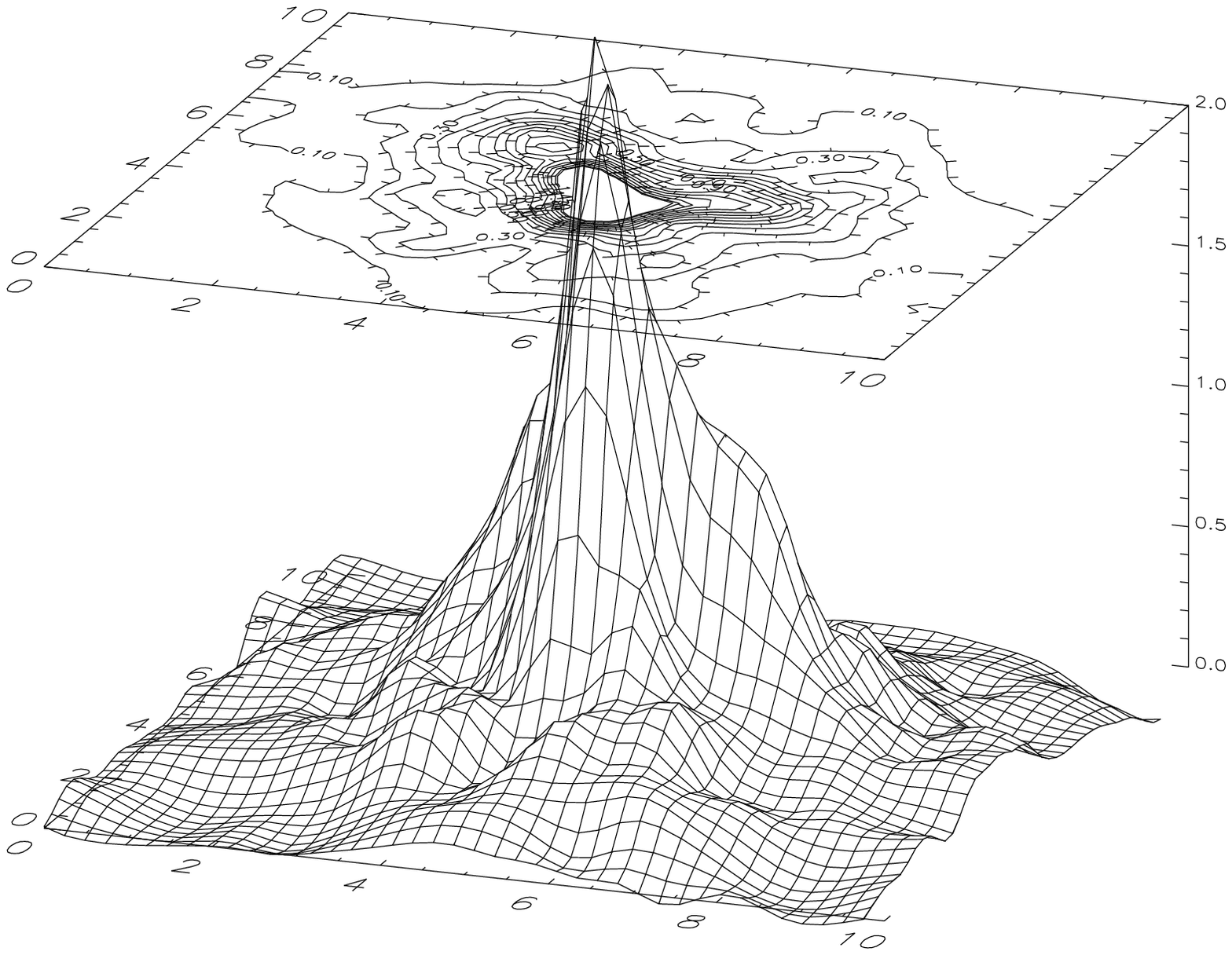}{10}

\xfigure{6}{The mean mass density $\bar \kappa$ 
averaged over the data field
$\cal U$ as a function of the assumed redshift distribution of the
sources. The mean mass density is obtained such that the minimum of
the locally averaged mass density is zero.  For redshift distributions
with $\beta=1$ we use crosses, for $\beta=1.5$ triangles and for
$\beta=3$ dashes. The left panel shows the mean mass density 
obtained from the
reconstruction as a function of the mean redshift $\ave z$, the right
panel as a function of $\ave w$ defined in (A2.1).  For producing the
synthetic galaxy images we use for their redshift distribution $\ave
z_{\rm true}=1$ and $\beta_{\rm true}=1$, i.e., $\ave w=0.44$. The
true mean mass density of the cluster is $\bar
\kappa_{\rm true}=0.12$}{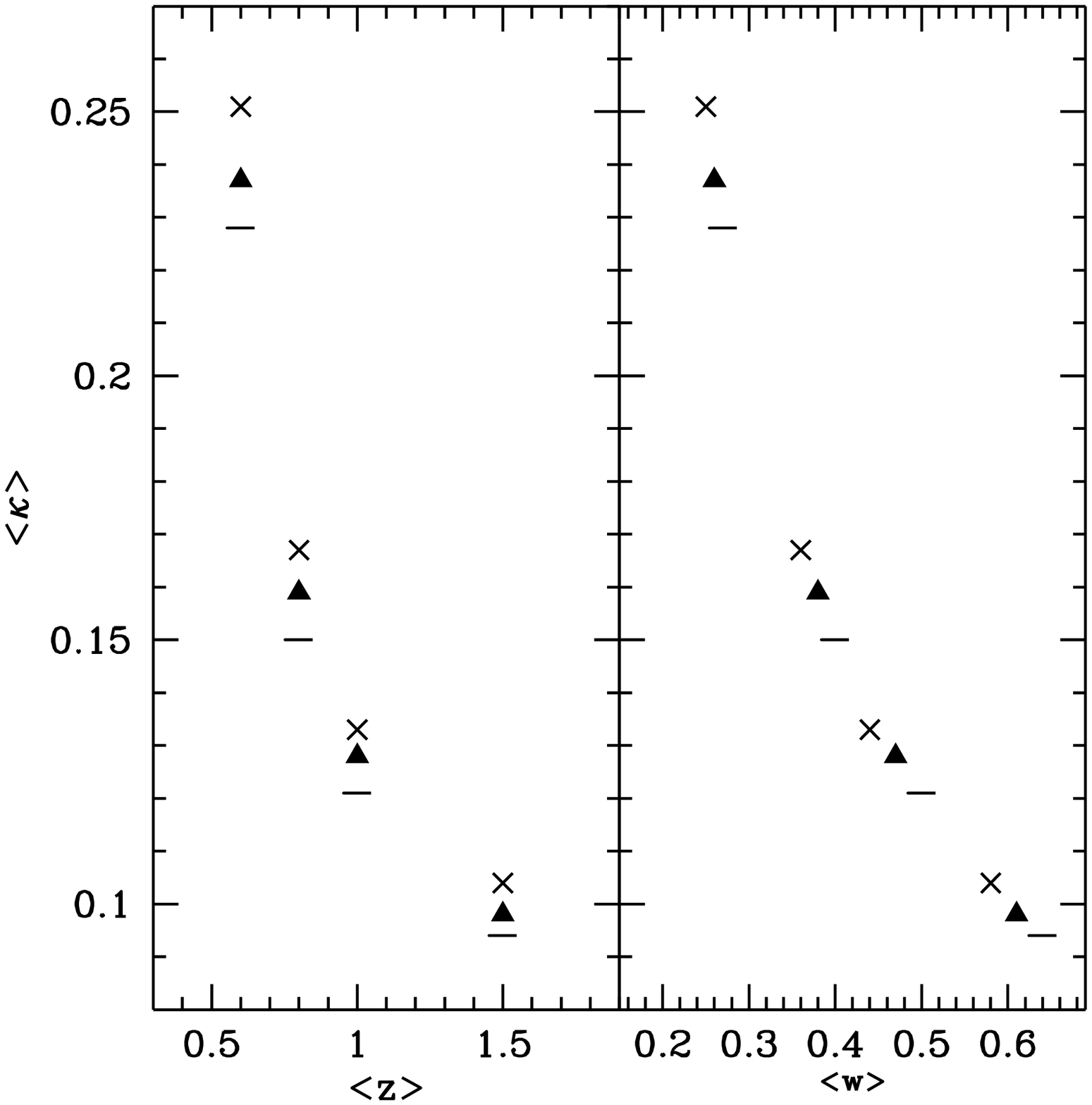}{7}

\subs{6.2 Breaking the mass invariance in practice}
We now use Eq.\ts(5.12) to break the mass invariance.  
First we choose the parameter $\alpha$ describing the number
counts versus flux of the faint galaxies. Again, the source
ellipticity- and redshift distribution is given through (3.9) and
(6.1) with parameters as given in Table\ts 1. The spatial 
distribution of the
galaxy images is given through Eq.\ts(5.5), where
$\ave{\mu^{\alpha-1}(\vc \theta,z)}_z$ is calculated from integrating
the `true' local magnification weighted with the redshift distribution
of the sources. We keep the number of galaxy images (lens plane) fixed and
 use $N=5260$ galaxy images for our simulations.

Next we perform the mass reconstruction according to Eq\ts.(4.9) for
different values of $\bar \kappa$. From the resulting mass and shear
map we calculate the map $\ave{\mu^{\alpha-1}(\vc \theta,z,\bar \kappa)}_z$,
where we assume to know the true redshift distribution of the sources
and the true slope $\alpha$.
Finally, we derive from Eq.\ts(5.12) the number of expected galaxies
for the assumed value of $\bar \kappa$ and calculate from that with
Eq.\ts(5.13) the function $\chi^2(\bar \kappa)$. 
The resulting function $\chi^2(\bar \kappa)$ is shown in Fig.\ts 7 for
different values of the parameter $\alpha$.

\xfigure{7}{The $\chi^2(\bar \kappa)$ as a function of the assumed
value of the mean mass density $\bar \kappa$ used for the mass
reconstruction. We use different parameters $\alpha$ to describe the
dependence of the number counts on the flux ($F(>S)\propto
S^{-\alpha}$) of the faint galaxies. We assume to know the redshift
distribution of the sources and the parameter $\alpha$. The true
values of the mean mass density is $\bar\kappa_{\rm
true}=0.12$}{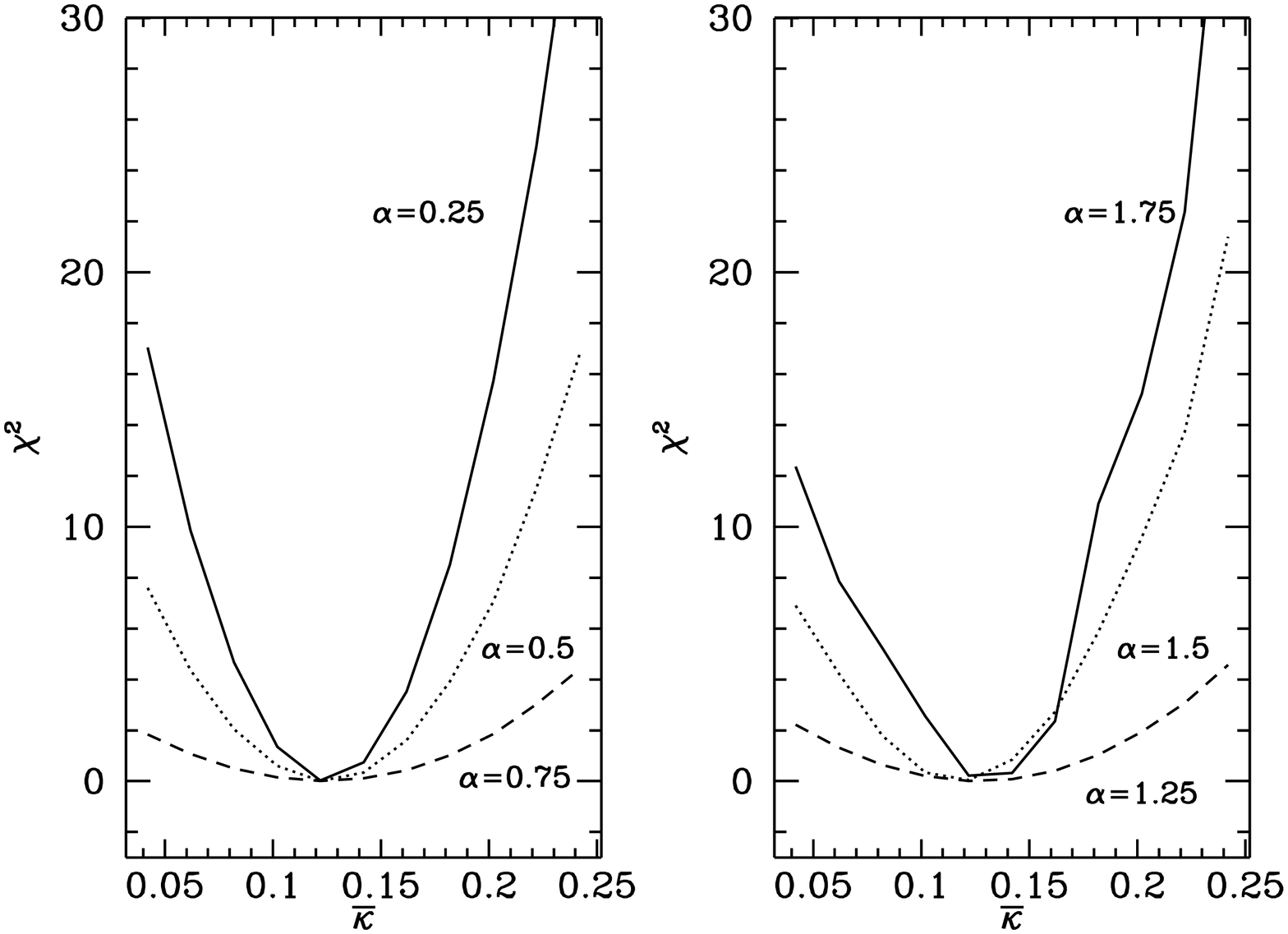}{10}

We find that all curves $\chi^2(\bar \kappa)$ show a minimum close to
$\bar \kappa_{\rm true}=0.12$, but the minima are narrower the more
the slope $\alpha$ deviates from one. This is because the larger
$\abs{1-\alpha}$, the stronger the magnification bias (or anti-bias)
and the stronger the dependence of $\ave N(\bar \kappa)$ is on $\bar
\kappa$. Hence, for the same number density of galaxy images, the
significance of rejecting values of $\bar \kappa$ increases with
increasing $\abs{1-\alpha}$. From our simulations we find that the
mean surface mass density is $0.09\le \bar \kappa \le 0.16$ ($2\sigma$
level) for $\alpha=0.25$, $0.06 \le \bar \kappa \le 0.18$ for
$\alpha=0.5$ and $0 \le \bar
\kappa \le 0.23$ for
$\alpha=0.75$. The true value of the mean surface mass density is
$\bar \kappa_{\rm true}=0.12$. Increasing the number density of the
observed galaxy images improves the significance levels, because the
$\chi^2$ function is roughly proportional to the number of observed
galaxies. We note that the galaxies used for determining $\bar \kappa$
can be different from that used for the mass reconstruction, since the
shape of the galaxies has not to be measured for the former purpose,
increasing the number density of available galaxy images considerably.
A number density of $n_{\rm gal}=100/{\rm arcmin}^2$ would reduce the
$2\sigma$ confidence levels to $(0.08,0.17)$ for $\alpha=0.5$ and to
$(0.10,0.15)$ for $\alpha=0.25$.

In practice, the redshift distribution of the sources is poorly
known and the mean redshift $\ave z$ can be over- or underestimated in
a cluster mass reconstruction. If the true mean redshift is $\ave z
_{\rm true}=1$ and if we assume for the reconstruction $\ave z=0.8 $
[1.5], then we obtain the $2\sigma$ confidence limits $\bar
\kappa\in(0.11,0.23)$ [$\bar \kappa \in(0.08,0.16)$].  To derive these
limits we used a slope $\alpha=0.5$ for the number counts of the faint
galaxies. 

\sec{7 Discussion}
The present paper should be considered as a further step in
building up the ground for nonlinear cluster reconstruction
techniques. In contrast to Papers\ts I\ts\&\ts II we have taken into
account a redshift distribution of the sources.  Our main results can
be summarized as follows:

(1) We have related the local expectation values of the moments $\eps^n$ of
the image ellipticities to the local lens parameters $\gamma$ and
$\kappa$ in Eq.\ts(3.6). These moments depend in a complicated way on
the redshift distribution of the sources as shown in
Eqs.\ts(3.6) and (3.7). In Sect.\ts4.3 we have shown that in the case of
non-critical clusters the local expectation values $\ave{\eps ^n}$
depend only on the local lens parameters $\gamma$ and $\kappa$ and on
few moments $\ave {w^k}$ of the effective distance $w$ of the sources
from the lens, weighted with the redshift distribution of the sources
[see Eqs.\ts(4.12), (4.15), and (4.16)].

(2) In Sect.\ts 4.2, we have generalized 
the inversion method developed in
Paper II such that the redshift distribution of the sources is
accounted for.  This is important if the cluster is at a high
redshift ($z_\d\gtrsim 0.2$), because then the effective lensing
strength for sources with $z_{\rm s}\approx 0.5$ is significantly 
different from
that for sources with $z_{\rm s}\approx 1$.  For the reconstruction
the redshift distribution of the sources has to be assumed and the
inversion equation is solved iteratively.  In Sect.\ts6 we have
applied the reconstruction technique to synthetic data and shown
that the reconstructed mass density is in good agreement with the
original one. Compared to the mass reconstruction in the case of a
single source redshift, the noise is slightly increased if the same
number density of galaxy images is used.

(3) For non-critical clusters we have simplified the inversion method in
Sect.\ts4.3 such that only one integration is necessary. Then the
inversion equation is very similar to that developed previously for a
single redshift of the sources (see Sect.\ts4.1). For the
reconstruction of a non-critical cluster, one only has to assume the
first two moments $\ave w$ and $\ave{w^2}$ of the effective distance
$w$ weighted with the redshift distribution. The weaker the cluster,
the less information on the redshift distribution has to be available
for the reconstruction of the mass distribution. For a weak
cluster it is sufficient to know the first moment $\ave w$.

(4) We have shown that the mean surface mass density $\bar \kappa$ is still
a free variable in the inversion equation (4.9), i.e., there still
remains a global invariance transformation for the surface mass
density field which leaves the field of $\ave{\eps}$ unchanged. For
non-critical clusters, this invariance transformation is given
explicitly in Eq.\ts(4.20) and it is similar to that derived
previously for a
single redshift of the sources [see Eq.\ts(4.5)]. 

(5) In Sect.\ts5 we have discussed some ideas to break the mass
invariance: maximizing the likelihood for the observed image
ellipticities, using the second moments $\ave{\eps^2}$ of image
ellipticities or the magnification effect on the position of the
galaxy images. For all these cases, the mass invariance is broken only
in the presence of a redshift distribution of the sources.  We find
that the invariance is broken too weakly to make use of it in
practice.

(6) Using the magnification effect on the number density of the galaxy
images (see Broadhurst, Taylor \& Peacock 1995) is the most promising
way to break the mass degeneracy. In contrast to Broadhurst et
al.\ts (1995) and Broadhurst (1995), we do not use the observed local
number density to derive the 
local surface mass density (because this local estimate  
has a considerably larger noise) but we use the total number of
observed images to determine the mean mass density: 
we assume a value for $\bar \kappa$, perform the mass
reconstruction, then calculate the local magnification and from that
the number of expected galaxy images in the field. Comparing this with
the number of `observed' galaxy images via a $\chi^2$-analysis gives a
confidence interval for $\bar \kappa$. This method works only if the
distribution $F(>S)\propto S^{-\alpha}$ of the unlensed sources with
flux larger than $S$ has a slope $\alpha \ne 1$, because for
$\alpha=1$ the number density of lensed objects is not changed
compared to the unlensed one.  The confidence interval obtained for
$\bar \kappa$ is the narrower, the larger $\abs{\alpha-1}$ is, because
the magnification bias (anti-bias) increases with increasing
$\alpha>1$ (decreasing $\alpha<1$).

(7) In Sect.\ts 6 we have used the same numerically modelled cluster as in
Paper II for testing the reconstruction technique. We reconstruct the
mass density for the case that (i) the assumed redshift distribution
is the true one, (ii) the mean redshift is overestimated and (iii)
underestimated. The mean mass density $\bar \kappa$ is adjusted such
that the minimum of the reconstructed mass density is zero. We find
that the substructure of the cluster is nicely recovered regardless of
the assumed redshift distribution. The total mass detected is within
$10\%$ of the true mass for (i), underestimated for (ii) and
overestimated for (iii).  Using the method as describe above in (6), we
find that reliable limits on the mean mass density can be derived with
a significance increasing with increasing number density of galaxy
images and increasing $\abs{\alpha-1}$. We note that for a successful
application of this method on real data it is essential (a) to know
the number density of the unlensed sources which would have been
detected in the absence of lensing but under the {\it same observing
conditions}, (b) to choose a colour which gives an appropriate slope of
the faint galaxies in flux ($\abs{ \alpha-1} $ large !) and (c) to
know the mean redshift of the sources considered.

Finally, we should mention that although the cluster reconstruction
method developed in this and our earlier papers is considerably more
complicated than straight application of the original Kaiser \&
Squires (1993) method, these modifications are essential if applied to 
WFPC2 observations of a cluster center. For the one case we
considered, i.e., the cluster Cl 0939+4713 (Seitz et al.\ts 1996),
the small field-of-view of the WFPC2, the lensing strength of the
cluster center, and the fairly high redshift $z_{\rm d}=0.4$ of the
cluster make it absolutely necessary to apply an unbiased finite-field
inversion technique, to use a non-linear reconstruction method, and to
account for a redshift distribution of the galaxies.

{\it Acknowledgement:}
\def\SFB{{We thank Achim Wei\ss\ for carefully reading, and very
useful comments on our 
manuscript. This work was supported by the ``Sonderforschungsbereich
375-95 f\"ur
Astro--Teil\-chen\-phy\-sik" der Deutschen For\-schungs\-ge\-mein\-schaft.}}
\SFB

\sec{Appendix\ts 1}
Let us derive Eq.\ts(3.5), first considering the case $\abs{g(z)}\le
1$. We 
define $\epsilon_{\rm s}=y\exp\rund{2\i\varphi}$ and
$u=\exp\rund{2i\varphi}$, so that $\d\varphi= -i\,\d u/u$, 
use Eq.\ts(3.4b) and
begin with the $\varphi$-integration:
$$
\eqalign{
\int_0^{2\pi}\d\varphi\;  \epsilon^n(\epsilon_{\rm s},z)
=&\int_0^{2\pi}\d\varphi\; \rund{\epsilon_{\rm s}-g(z)\over 1-g^*(z)\epsilon_{\rm s}}^n
=-\i \oint \d u \rund{yu-g(z)\over 1-g^*(z)yu}^n{1\over u}\cr
=&2\pi \eck{-g(z)}^n\; .\cr}
\eqno(A1.1)
$$
In the last step, we used that the integrand has one pole at $u=0$
where the
corresponding residuum is $\eck{-g(z)}^n$; the other pole at
$u=1/(g^* y)$ lies outside the unit circle and thus does not
contribute. As a result we obtain for  $\abs{g(z)}\le 1$, 
$$
\ave{\epsilon^n}_{\epsilon_{\rm s}}(z)=\int_0^1\d y \; y
p_{\epsilon_{\rm s}}(y) \eck{-g(z)}^n=\eck{-g(z)}^n
\ . 
\eqno(A1.2)
$$
Similarily, we derive for $\abs{g(z)}>1$:
$$
\eqalign{
\int_0^{2\pi}\d\varphi\;  \epsilon^n(\epsilon_{\rm s},z)
&=\eck{\int_0^{2\pi}\d\varphi\;  \eck{\epsilon^*(\epsilon_{\rm s},z)}^n}^*
=\eck{ \int_0^{2\pi}\d\varphi\; \rund{1-g^*(z)\epsilon_{\rm s}\over \epsilon_{\rm s}-
g(z)}^n }^* \cr
&=\eck{ -\i \oint \d u \rund{1-g^*(z)uy\over yu-g(z)}^n{1\over u} }^* 
=2\pi \rund{-1\over g^*(z)}^n\cr}
\ , \eqno(A1.3)
$$
since the pole at $u=g/y$ lies again outside the unit circle; we thus
find that for  $\abs{g(z)}>1 $, 
$$
\ave{\epsilon^n}_{\epsilon_{\rm s}}(z)=\int_0^1\d y \; y
p_{\epsilon_{\rm s}}(y) \rund{-1\over g^*(z)}^n=\rund{-1\over g^*(z)}^n
\ . 
\eqno(A1.4)
$$

\sec{Appendix\ts 2}
Here we derive approximations for $X_1(\kappa)$ and $X_2(\kappa)$ for
the case of non-critical clusters. 
This non-criticality implies that
$1-w \kappa>0$. We can expand the integral in (4.10),
using
$$
\ave{w^n}:=\int_0^1  \d w\; p_w(w) w^n=\int_{z_{\rm d}}^\infty  \d z\;
p_z(z) \; w^n(z)\ ,
\eqno(A2.1)
$$
to obtain
$$
X_1(\kappa)=\int_0^1 \d w\; p_w(w){-w\over 1-\kappa
w}=-\int_0^1  \d w\;p_w(w)w \sum_{k=0}^\infty (\kappa w)^k =
-\sum_{k=0}^\infty \kappa^k\ave{w^{k+1}} \ .
\eqno(A2.2)
$$
It is easily seen from (4.10) that for $n\ge 1$ the following relation holds:
$$
X_{n+1}(\kappa)=- {1\over n}{\d X_n(\kappa)\over \d \kappa}\ .
\eqno (A2.3)
$$
As it turns out, whereas the power series in (A2.2) converges, a
Pad\'e approximation behaves much better. From (A2.2) and (A2.3), we
obtain for the lowest-order approximation (see
e.g. Press et al.\ts 1992, page 194 ff.)
$$
\eqalign{ X_1(\kappa) &\approx {-\ave w\over 1-\kappa {\ave{w^2} \over
\ave w}} \ ,\cr X_2(\kappa) &\approx {\ave{w^2}\over \rund{1-\kappa
{\ave{w^2}\over
\ave w}}^2} \ , \cr}
\eqno(A2.4)
$$
and for the next higher order,
$$
\eqalign{
X_1(\kappa) &\approx -{\ave w+\kappa\eck{\ave{w^2}-{\ave{w^3}\ave w\over
\ave{w^2}}}
\over 1-\kappa {\ave{w^3}\over \ave{w^2}}} \ , \cr
X_2(\kappa)&\approx {\ave{w^2}\over \rund{1-\kappa{\ave{w^3}\over
\ave{w^2}}}^2} \ . \cr}
\eqno(A2.5)
$$
\xfigure{8}{Comparing different approximations for $X_1(\kappa)$ (left panel)
and $X_2(\kappa)$ (right panel) as defined in Eq.(3.7) as a function
of $\kappa$ for the redshift distribution $p_z(z)$ shown in Fig.\ts1.
The dotted lines show the approximations given in Eq.\ts(A2.4), the
dashed one those of Eq.\ts(A2.5) and the long dashed line the
expansion given in Eq.\ts(A2.2) up to $k=2$ as an approximation for
$X_1(\kappa)$, and its first derivative as an approximation for
$X_2(\kappa)$ ($X_2=-{\partial X_1/\partial \kappa}$), respectively. The
solid lines show the exact result for $X_1$ and ${X_2}$
obtained from performing the integration (4.10) }
{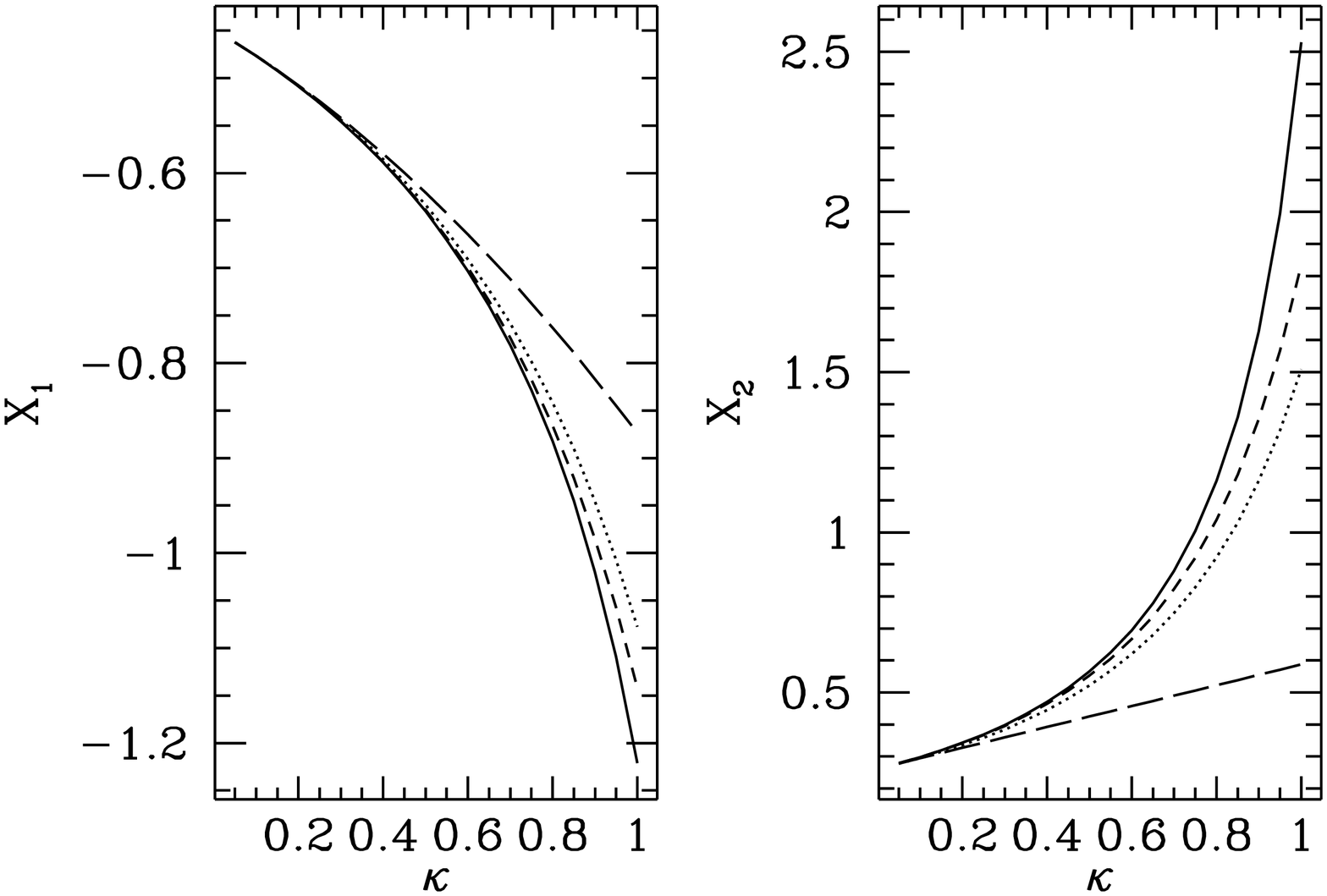}{9}

In Fig.\ts 8, left panel, we compare $X_1(\kappa)$ with the
approximations (A2.4) and (A2.5), and the expansion (A2.2) up to the term
$k=2$ for a cluster with $z_{\rm d}=0.4$ and the redshift distribution
shown in Fig.\ts 1 (i.e., for which $\ave z= 1$).  We find that the
approximations (A2.4) and (A2.5) are excellent for $\kappa \lesssim 0.6$,
good for $0.6\lesssim
\kappa  \lesssim 0.8$, and clearly deteriorate for
$\kappa \gtrsim 0.8$. As expected, approximation (A2.5) is always
better than (A2.4). In Fig.\ts 8, right panel, we compare
$X_2(\kappa)$ with the approximations given above. We
find qualitatively the same result, but the deviation of $X_2(\kappa)$
from the approximations (A2.4) and (A2.5) becomes important for smaller
values of $\kappa$, $\kappa\gtrsim 0.6$, say.  We note that these
approximations become even better if the difference between
mean redshift of the sources and the cluster redshift increases.

As a result, we obtain that the approximation (A2.4) for $X_1(\kappa)$
is sufficiently accurate over
a wide range of $\kappa$ (if $\kappa$ approaches values close to 0.8,
the cluster most likely is critical anyway). 
Therefore, one can use
approximation (A2.4) for (most) non-critical clusters.

\sec{Appendix\ts 3}
In the case of a non-critical cluster (or at positions of a critical
cluster where $\kappa+\abs{\gamma}<1$) the ratio
$$
R:=\abs{ \ave{\eps^2}\over \ave{\eps}^2}\quad 
\eqno (A3.1)
$$
becomes according (4.10)
$$
R={X_2(\kappa)/ X_1^2(\kappa)}\quad .
\eqno (A3.2)
$$
Therefore, for a fixed redshift distribution of the galaxies, the
ratio $R$ depends on $\kappa$ only. If we identify
the expectation values again with the local averages,
$$
R\approx \abs{\bar{\eps^2}\over (\bar\eps)^2} \quad ,
\eqno (A3.3)
$$
a comparison of (A3.2) with (A3.3) can in principle determine the local
value of $\kappa$. In Fig.\ts 9 we show $R$ for a non-critical
cluster as a function of $\kappa$ (solid line) for  the redshift
distribution shown in Fig.\ts 1 ($\beta=1=\ave z$).
The following approximations for $R(\kappa)$ can be derived from
(4.11) and the relations in Appendix 2:
$$
 R(\kappa)\approx
\cases{ 
\displaystyle
{\ave{w^2}\over \ave w^2}=f= R(0)\    & \hbox{for} 
$\kappa\to 0$ , \phantom{.0}\qquad \phantom{123456789101112}  (A3.4a) \cr
\displaystyle
f\; 
{\eck{1+\kappa\rund{{\ave{w^2}\over \ave{w}}-{\ave{w^3}
\over \ave{w^2}}}}^{-2}} & \hbox{for}   $ \kappa\lesssim 0.2$ ,
\qquad\phantom{123456789101112}   (A3.4b)  \cr
\displaystyle
 f \; { \rund{1-\kappa {\ave{w^2}\over \ave w}}^2
\over \rund{1-\kappa {\ave {w^3}\over \ave{w^2}} }^2}  &
\hbox{for} $\kappa \lesssim 0.8 $ . \qquad\phantom{123456789101112} (A3.4c)\cr 
}
$$
We obtain Eq.\ts(A3.4a) using the approximations (A2.4) for $X_1(\kappa)$
and $X_2(\kappa)$. For Eq.\ts(A3.4b) we use the approximations (A2.5)
for $X_1$ and $X_2$, and in the case of Eq.\ts(A3.4c) we use the
approximation (A2.5) for $X_2$ and the approximation (A2.4) for
$X_1$. We compare these approximations with $R=X_2(\kappa)/X_1^2(\kappa)$
in Fig.\ts 9: for the chosen redshift distribution, (A3.4a) gives
$R\approx 1.29$, approximation (A3.4b) is shown as the dotted line and
is obviously a good approximation for small $\kappa \lesssim 0.2$,
whereas approximation (A3.4c) is shown as the dashed line and fits
well up to
$\kappa \lesssim 0.8$. For higher surface mass densities the cluster
is most probably critical and Eq.\ts(A3.2) does not hold anyway, but
has to be replaced according to Eq.\ts(3.6); then, $R$ becomes a
function of $\gamma$ and $\kappa$.  Eq.\ts(A3.4c) can be
inverted to obtain $\kappa$ as a function of the observable quantity
$R$,
$$
\kappa\approx {\sqrt{R\over f}-1\over {{ \ave{w^3}\over \ave{w^2}}
\sqrt{R\over f}-{ \ave {w^2}\over \ave w}}} \ .
\eqno(A3.5)
$$
In Fig.\ts 9 (right panel) we compare the surface mass density
$\kappa$ derived from Eq.\ts(A3.5) with the true one and find indeed a
good agreement for $\kappa \lesssim 0.8$. 

\xfigure{9}{The solid line (left panel) shows the ratio
$R=\abs{\ave{\eps^2}}/\abs{\ave \eps ^2}$ as a function of the surface
mass density $\kappa$ in the case of a non-critical cluster, for which
$R=X_2(\kappa)/X_1^2(\kappa)$. The dotted line shows the approximation
(A3.4b) and the dashed line (A3.4c). We can invert (A3.4c) to
derive the approximation (A3.5) for $\kappa$. The resulting values
$\kappa^*$ are compared to the true one $\kappa$ in the right
panel}{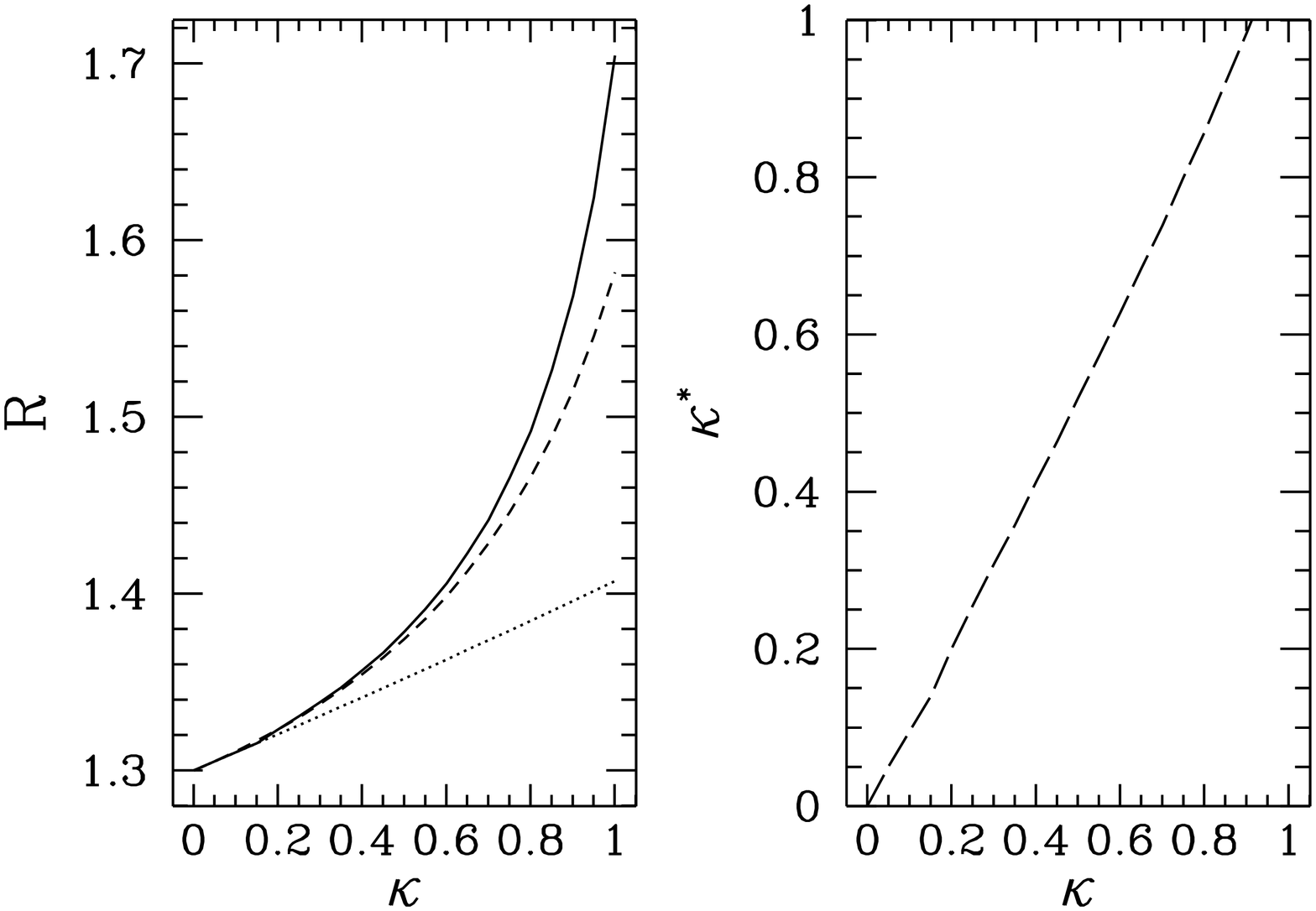}{9}

From Eq.\ts (A3.5) one might conclude that $\kappa(\vc \theta)$ can be
directly obtained from observing $R(\vc \theta)$ without using any of
the inversion techniques derived in the past. Unfortunately this is
not the case because $R(\vc \theta)$ can not be determined precisely enough
from local observations. To obtain a reliable estimate
for a local surface mass density $\kappa \gtrsim 0.4$,
one would need about $1000$ galaxy images.

Thus, we try to derive the mean surface mass density $\bar \kappa$ by
measuring $R(\vc \theta)$: assuming a value for $\bar \kappa$ we perform
the reconstruction according to (4.9) and derive from the mass and
shear map the corresponding map of $\tilde R(\vc \theta)$, which then
depends on the assumed value for the mean mass density $\bar
\kappa$. Finally, we compare this with the measured $R(\vc
\theta)$, i.e., we search for $\bar \kappa$ which minimizes
$\chi^2=\ave{\rund{\tilde R(\vc \theta)-R(\vc \theta)}^2}_{\cal U}$,
averaged over the data-field $\cal U$. Again, we find that the mean
surface mass density can not be determined in that way, because the
local mean image ellipticities $\ave{\eps}$ and $\ave{\eps^2}$ can not
be derived with sufficient accuracy from a reasonable
number density of galaxy images.

We note that we tried to determine $\bar \kappa$ in various other ways 
from moments of image ellipticities: assuming a value for $\bar
\kappa$ we performed the reconstruction and calculated from the mass and
shear map the local expectation values of $\ave{ \eps^n}(\bar \kappa)$
according to Eq.\ts(3.6). Then, we compared that with the `measured'
local means $\bar{\eps^n}$ and minimized $\chi^2=\abs{\ave{
\eps^n}(\bar \kappa)- \bar{\eps^n}}^2_{\cal U}$ varying the value 
$\bar \kappa$ used for the reconstruction. All these attempts failed
for resonable assumptions for the number density of galaxy images,
showing that the image ellipticities provide not enough information to
break the mass degeneracy in practice.

\sec{References}

\def\ref#1{\vskip5pt\noindent\hangindent=40pt\hangafter=1 {#1}\par}
\ref{Bartelmann, M. \ 1995, A\&A, 303, 643.}
\ref{Bartelmann, M. \& Narayan, R. \ 1995, ApJ, 451, 60.}
\ref{Bartelmann, M., Narayan, R., Seitz, S. \& Schneider, P.\ 1996,
ApJ Letters, submitted.}
\ref{Broadhurst, T.J., Taylor, A.N. \& Peacock, J.A. \ 1995, ApJ 438,
49.}
\ref{Broadhurst, T.G. \ 1995, preprint astro-ph/9505013.}
\ref{Brainerd, T., Blandford, R. \& Smail, I. 1995, preprint astro-ph/9503073.}
\ref{Infante, L. \& Pritchet, C.J. 1995, preprint.}
\ref{Kaiser, N. \& Squires, G.\ 1993, ApJ 404, 441 (KS).}
\ref{Kaiser, N. 1995, ApJ 493, L1.}
\ref{Kaiser,N., Squires, G., Fahlman, G.G., Woods, D. \& Braodhurst,
T. \ 1994, preprint astro-ph/9411029.}
\ref{Kneib, J.P., Ellis, R.S., Smail, I.R., Couch, W.J., Sharples, R., 1995,
ApJ submitted.}
\ref{Kochanek, C.S.\ 1990, MNRAS 247, 135.} 
\ref{Press, W.H., Teukolsky, S.A., Vetterlin, W.T. \& Flannery,
B.P.\ 1992, {\it Numerical Recipes}, Cambridge University Press.}
\ref{Miralda-Escude, J.\ 1991, ApJ 370, 1.}
\ref{Schneider, P. 1995, A\&A 302, 639.}
\ref{Schneider, P. \& Seitz, C.\ 1995, A\&A 294, 411 (Paper\ts I).}
\ref{Seitz, C. \& Schneider, P.\ 1995, A\&A 297, 287 (Paper\ts II).}
\ref{Seitz, C., Kneib, J.P., Schneider, P. \& Seitz, S. 1995, A\&A, submitted.}
\ref{Seitz, S. \& Schneider, P. 1996, A\&A 305, 383.}
\ref{Smail, I., Hogg, D.W., Yan, L. \& Cohen, J.G. \ 1995b, ApJ, 449, L105.}
\ref{Tyson, J.A., Valdes, F. \& Wenk, R.A. 1990, ApJ 349, L1.}

\end